\documentclass[12pt]{article}
\usepackage{epsfig}
\usepackage{amsfonts}
\usepackage{latexsym}
\usepackage{amsmath,amssymb}
\usepackage{mathrsfs}
\usepackage{hyperref}
\usepackage{setspace}
\usepackage{color}
\usepackage{bm}
\usepackage{slashed}
\numberwithin{equation}{section}

\begin{document}

\begin{titlepage}
\unitlength = 1mm
\begin{flushright}
OU-HET-1017
\end{flushright}

\vskip 1cm
\begin{center}

{\Large {\textsc{\textbf{Nonclassical primordial gravitational waves \\\vskip 1mm
from the initial entangled state}}}}

\vspace{1.8cm}
Sugumi Kanno

\vspace{1cm}

\shortstack[l]
{\it $^*$ Department of Physics, Osaka University, Toyonaka 560-0043, Japan 
}

\vskip 1.5cm

\begin{abstract}
\baselineskip=6mm
The nonclassicality of primordial gravitational waves (PGWs) is characterized in terms of sub-Poissonian graviton statistics. The sub-Poissonian statistics are realized when quantum states are squeezed coherent states. In the presence of matter fields, the Universe experiences the squeezed coherent state during inflation. The condition to realize the sub-Poissonian graviton statistics is translated into the frequency range of gravitational waves. If the initial state is the Bunch-Davies vacuum, there is another necessary condition between phases of squeezing and coherent parameters. Here, we extend the initial state to entangled states. We consider $\alpha$-vacua as the initial entangled state that are more general de Sitter invariant vacua than the Bunch-Davies vacuum.  We find that, unlike the Bunch-Davies vacuum, PGWs generated in the initial entangled state become sub-Poissonian without requiring the condition between the phases.

\end{abstract}

\vspace{1.0cm}

\end{center}
\end{titlepage}

\pagestyle{plain}
\setcounter{page}{1}
\newcounter{bean}
\baselineskip18pt

\setcounter{tocdepth}{2}

\tableofcontents

\section{Introduction}
\label{section1}

One of the greatest achievements of inflationary cosmology is that the connection between the quantum theory of the microscopic world and the large scale structure of the macroscopic world. The idea that the Universe has a quantum mechanical origin is now one of the cornerstones of inflationary cosmology. Primordial gravitational waves (PGWs) also arise out of original minute quantum fluctuations during inflation. However, any compelling observational evidence for the quantum nature of the initial fluctuations has not yet found. 

Quantum entanglement is an essential feature of quantum physics that correlations are shared between distant particles even beyond the cosmological horizon~\cite{Einstein:1935rr}. Recently, it was shown that quantum fields of causally disconnected regions in de Sitter space is entangled~\cite{Maldacena:2012xp, Kanno:2014lma, Iizuka:2014rua, Kanno:2014bma, Kanno:2016gas, Kanno:2016qcc, Choudhury:2017bou, Choudhury:2017qyl, Albrecht:2018prr}.  If we can find the observational evidence of the initial quantum fluctuations, we might be able to find the information about entanglement with other Universes encoded in them. 

The recent direct detection of gravitational waves in 2015~\cite{Abbott:2016blz} encourages us to challenge these problems. Currently, to detect PGWs is an important target for gravitational physics~\cite{Kawamura:2011zz, AmaroSeoane:2012km}. Since they interact very weekly with matter, travel through the Universe virtually unimpeded, they give us information about the original minute quantum fluctuations during inflation. Furthermore, if PGWs were detected, the detection could be regarded as a proof of inflationary cosmology. This is because the energy scale that generates PGWs has to be around GUT scale in order to detect them at present and it is difficult to find a possible scenario other than the inflationary scenario to realize such a high energy scale. On top of that, if we succeeded in detecting nonclassical PGWs, it would imply discovery of gravitons.

In this work, we characterize nonclassicality of PGWs in terms of sub-Poissonian graviton statistics as is known in quantum optics~\cite{Agarwal:2012}. The particle number distribution for coherent fields is Poissonian and any distribution which is wider than Poissonian is called super-Poissonian. Since the particle number distribution in classical theory is always super-Poissonian, it follows that sub-Poissonian distribution which is narrower than Poissonian must be a signature of nonclassicality. 

In our previous work~\cite{Kanno:2018cuk}, we studied graviton statistics of the inflationary Universe when the initial state is the Bunch-Davies vacuum. We found that the presence of matter fields during inflation makes graviton statistics sub-Poissonian. The condition to realize the sub-Poissonian graviton statistics is translated into the frequency range of gravitational waves. We showed that PGWs with frequency higher than $10$ kHz enable us to observe their nonclassicality if the phases of parameters satisfy a necessary condition. In this work, we extend the initial state to entangled states. We consider $\alpha$-vacua as the initial entangled state that are more general de Sitter invariant vacua. We show that unlike the Bunch-Davies vacuum, PGWs generated in the initial entangled state become sub-Poissonian without requiring the condition between phases.

The organization of this paper is as follows. We start in section~\ref{section2}, by reviewing the regimes of graviton statistics and introduce the Fano factor, a useful measure to distinguish the regime of graviton statistics. In section~\ref{section3}, we introduce quantum states and find that squeezed coherent states produce sub-Poissonian statistics. In section~\ref{section4}, we review our previous paper~\cite{Kanno:2018cuk} that studied  nonclassical PGWs generated in the Bunch-Davies vacuum. In section~\ref{section5}, we calculate graviton statistics in the initial entangled state. We summarize our result and discuss the possible detection of the nonclassical PGWs with Hanbury Brown and Twiss interferometry in section~\ref{section6}. In appendix~\ref{appA}, we give some formulas used in computation in section~\ref{section5}, appendix~\ref{appB} gives short notes on useful relations between coherent and squeezing operators, and appendix~\ref{appC} contains the details of the result of graviton statistics in the initial entangled state.

\section{Graviton statistics and Fano factor}
\label{section2}

In this section, we characterize graviton statistics by counting graviton numbers in a given state. To do this, we see the probability of finding $n$ gravitons. As a useful measure to distinguish the regime of graviton statistics, we introduce the Fano factor $F$ defined by the ratio of the variance squared to the mean such as
\begin{eqnarray}
F=\frac{(\Delta n)^2}{\langle n\rangle}\,.
\label{fano}
\end{eqnarray}

If the variance is equal to the mean number $\Delta n =\langle n\rangle$, it is called Poisson distrubition. Then the Fano factor becomes
\begin{eqnarray}
F=1\,,\qquad{\rm for}\quad{\rm Poissonian}\,.
\end{eqnarray} 
If the distribution becomes wider than Poissonian, that is, $\Delta n >\langle n\rangle$, it is called super-Poissonian and the Fano factor is 
\begin{eqnarray}
\hspace{1.3cm} F>1\,\,,\qquad{\rm for}\quad{\rm super}{\rm-Poissonian}\,.
\end{eqnarray}
The point here is that any classical theory leads to super-Poissonian distribution and the Fano factor is above one. Therefore, any distribution narrower than Poissonian, which is called sub-Poissonian, $\Delta n <\langle n\rangle$ or the Fano factor is 
\begin{eqnarray}
\hspace{1.3cm} F<1\,\,,\qquad{\rm for}\quad{\rm sub}{\rm-Poissonian}\,,
\end{eqnarray}
must correspond to nonclassical fields.

\section{Quantum states}
\label{section3}

In this section, we consider what kind of states make sub-Poissonian distribution. We see coherent states, squeezed states and squeezed coherent states~\cite{Koh:2004ez, Kundu:2011sg} in the following. 

\subsection{Coherent states}
\label{section3.1}

The coherent states $|\xi\rangle$ is defined as
\begin{eqnarray}
\hat{b}\,|\xi\rangle=\xi\,|\xi\rangle\,,
\label{coherent}
\end{eqnarray}
where the coherent parameter is written as $\xi=|\xi| e^{i\theta}$.
Thus, the coherent state remains unchanged by the annihilation of a particle. The formal solution of the eigenvalue equation is given by
\begin{eqnarray}
|\xi\rangle=\exp\left(\xi\,\hat{b}^\dag-\xi^*\hat{b}\,\right)|0\rangle\equiv\hat{D}(\xi)|0\rangle\,,
\end{eqnarray}
where $\hat{D}$ is an unitary operator called displacement operator.
Then the mean and the variance of particles are calculated as
\begin{eqnarray}
\langle\xi|\hat{n}|\xi\rangle=|\xi|^2\,,\qquad
\left( \Delta n \right)^2=\langle\xi|\hat{n}^2|\xi\rangle-\langle\xi|\hat{n}|\xi\rangle^2
=|\xi|^2\,,\qquad
\hat{n}=\hat{b}^\dag\hat{b}\,.
\label{statistics1}
\end{eqnarray}
Then, Fano factor Eq.~(\ref{fano}) becomes $F=1$. We find that the coherent state gives Poisson distribution.

\subsection{Squeezed states}
\label{section3.2}

The definition of the squeezed states $|\zeta\rangle$ is
\begin{eqnarray}
|\zeta\rangle=\exp\left(\zeta^*\,\hat{c}\,\hat{c}-\zeta\,\hat{c}^\dag\hat{c}^\dag\right)|0\rangle\equiv\hat{S}(\zeta)|0\rangle\,,
\label{squeezed}
\end{eqnarray}
where $\zeta=re^{i\varphi}$ and $r$ is the squeezing parameter. $\hat{S}$ is an unitary operator named squeezing operator. The operator $\hat{c}$ is obtained by the Bogoliubov transformation,
\begin{eqnarray}
\hat{c}=\cosh r\,\hat{b}-e^{i\varphi}\sinh r\,\hat{b}^\dag\,,\qquad
\hat{b}|\zeta\rangle=0\,.
\end{eqnarray}
The mean and the variance of particles are
\begin{eqnarray}
\langle\zeta|\hat{n}|\zeta\rangle=\sinh^2r\,,\quad
\left( \Delta n \right)^2=\langle\zeta|\hat{n}^2|\zeta\rangle-\langle\zeta|\hat{n}|\zeta\rangle^2=2\sinh^4r+2\sinh^2r\,,\quad
\hat{n}=\hat{c}^\dag\hat{c}\,.
\label{statistics2}
\end{eqnarray}
Then the Fano factor becomes
\begin{eqnarray}
F=2\sinh^2r+2\sinh r>1\,.
\end{eqnarray}
Thus the particle statistics in the squeezed state becomes super-Poissonian.

\subsection{Squeezed coherent states}
\label{section3.3}

Lastly, let us see the squeezed coherent states. The squeezed coherent state is defined as
\begin{eqnarray}
|\zeta,\xi\rangle=\hat{S}(\zeta)\hat{D}(\xi)|0\rangle\,.
\end{eqnarray}
The particle statistics in this state is calculated as
\begin{eqnarray}
\hspace{-5mm}
\langle\zeta,\xi|\hat{n}|\zeta,\xi\rangle&=&|\xi|^2\left[e^{-2r}\cos^2\left(\theta-\frac{\varphi}{2}\right)+e^{2r}\sin^2\left(\theta-\frac{\varphi}{2}\right)\right]+\sinh^2r\,,\nonumber\\
\left( \Delta n \right)^2&=&|\xi|^2\left[e^{-2r}\cos^2\left(\theta-\frac{\varphi}{2}\right)+e^{2r}\sin^2\left(\theta-\frac{\varphi}{2}\right)\right]+2\sinh^4r+2\sinh^2r\,.
\end{eqnarray}
If we take the limit $r\rightarrow 0$, the above recovers Eq.~(\ref{statistics1}). In the limit $\xi\rightarrow 0$, the particle statistics become Eq.~(\ref{statistics2}). The Fano factor is
\begin{eqnarray}
F=\frac{|\xi|^2e^{-4r}+2\sinh^2r+2\sinh^4r}{|\xi|^2e^{-2r}+\sinh^2r}\,,
\end{eqnarray}
where for simplicity, we assumed $\theta-\varphi/2=0$. Now two parameters $\xi$ and $r$ come in the Fano factor, then particle statistics can become Poissonian, super-Poissonian and sub-Poissonian. If the Fano factor satisfies 
\begin{eqnarray}
F<1\quad\Longleftrightarrow\quad
|\xi|^2e^{-2r}+\sinh^2r>|\xi|^2e^{-4r}+2\sinh^2r+2\sinh^4r\,,
\end{eqnarray}
then the particle statistics become sub-Poissonian and we have a chance to observe the nonclassicality.

\section{Review of graviton statistics in the Bunch-Davies vacuum}
\label{section4}

In the previous section, we find that the squeezed coherent state gives sub-Poissonian distribution. In this section, we review that the Universe has experienced the squeezed coherent state in the past. To explain this, we first consider how PGWs are generated by quantum fluctuations.

\subsection{PGWs generated by quantum fluctuations}
\label{section4.1}

The gravitational waves $h_{ij}(\eta, x^i)$ is expressed by the tensor perturbations in the metric
\begin{eqnarray}
ds^2=a^2(\eta)\left[-d\eta^2+(\delta_{ij}+h_{ij}) dx^idx^j\,\right]\,,
\end{eqnarray}
where $\eta$ is the conformal time, $a(\eta)$ is the scale factor, $x^i$ are spatial coordinates, and $\delta_{ij}$ and $h_{ij}$ are the Kronecker delta and the tensor perturbations which satisfy $h_{ij}{}^{,j}=h^i{}_i=0$. The indices $(i,j)$ run from $1$ to $3$. In order to quantize the tensor field, we decompose the tensor field $h_{ij}(\eta,x^i)$ in terms of the Fourier modes as
\begin{eqnarray}
a(\eta) h_{ij}(\eta, x^i) = {\frac{\sqrt{2}}{M_{\rm pl}}}\frac{1}{\sqrt{V}}\sum_{\bm k}\sum_{A} \ h^A_{\bm{k}}(\eta)\,e^{i {\bm k} \cdot {\bm x}} \ p_{ij}^A(\bm k)  \,,
\label{fourier}
\end{eqnarray}
where $p^A_{ij}({\bm k})$ is the polarization tensor normalized as $p^{*A}_{ij} p^B_{ij} =2 \delta^{AB}$ and the index $A$ denotes the polarization modes, 
for example, for circular polarization modes  $A=\pm$ and for linear polarization modes $A=+,\times$. Notice that we consider finite volume $V=L_{x }L_{y}L_{z}$ and 
 discretize the ${\bm k}$-mode with a width ${\bm k} = \left(2\pi  n_x/L_x\,,2\pi  n_y/L_y\,,2\pi  n_z/L_z\right)$ where ${\bm n}$ 
are integers in order to discuss graviton number distribution later.

In quantum field theory, the tensor field on the right hand side, $h^A_{\bm{k}}(\eta)$, is promoted to the operator. The operator $h^A_{\bm{k}}(\eta)$ satisfies
\begin{eqnarray}
h_{\bm k}^{\prime\prime A}+\left(k^2-\frac{a''}{a}\right)  h^A_{\bm k}=0\,.
\label{eom}
\end{eqnarray}
where $k$ is the magnitude of the wave number ${\bm k}$. In order to solve this, we need to determine the scale factor $a(\eta)$. As the Universe evolves, the scale factor changes as
\begin{eqnarray}
a(\eta)=\left\{
\begin{array}{l}
\vspace{0.2cm}
-\frac{1}{H \left( \eta -2\eta_1 \right)}\,,\hspace{1.1cm} {\rm for~(I)}\quad -\infty<\eta<\eta_1\,,\\
\vspace{0.2cm}
\frac{\eta}{H\eta_1^2}\,,\hspace{2.2cm} {\rm for~(R)}\quad \eta_1<\eta  \,,
\end{array}
\right.
\end{eqnarray}
where we assumed the Universe goes through an instantaneous transition from the inflationary epoch approximated by de Sitter space (I) to a radiation-dominated era (R) and the transition occurs at $\eta=\eta_1>0$. Then Eq.~(\ref{eom}) gives the positive frequency mode in each epoch as
\begin{eqnarray}
\left\{
\begin{array}{l}
\vspace{0.2cm}
v_{k}^{\rm I}(\eta)\equiv\frac{1}{\sqrt{2k}}\left(1-\frac{i}{k \left( \eta -2\eta_1 \right)}\right)e^{-ik \left( \eta -2\eta_1 \right)}\,,\\
\vspace{0.2cm}
v_{k}^{\rm R}(\eta)\equiv\frac{1}{\sqrt{2k}}\,e^{-ik \eta } \,.
\end{array}
\right.
\label{modefunction}
\end{eqnarray}

In the inflationary era, the operator $h^A_{\bm{k}}(\eta)$ is expanded as
\begin{eqnarray}
h^A_{\bm k}(\eta)=b^A_{\bm k}\,v^{\rm I}_k(\eta)+b_{-\bm k}^{A \dag}\,v_k^{\rm I *}(\eta)\,,\qquad
\left[b^A_{\bm k} , b_{\bm p}^{B\dag} \right]= \delta^{AB} \delta_{\bm k,\bm p}\,,
\label{pfm-I}
\end{eqnarray}
where $*$ denotes complex conjugation.  The operator $h^A_{\bm{k}}(\eta)$ should be the same even if we expand it by $v_{k}^{\rm R}(\eta)$ such as
\begin{eqnarray}
h^A_{\bm k}(\eta)=c^A_{\bm k}\,v^{\rm R}_k(\eta)+c_{-\bm k}^{A \dag}\,v_k^{\rm R *}(\eta)
\,,\qquad
\left[c^A_{\bm k} , c_{\bm p}^{B\dag} \right]= \delta^{AB} \delta_{\bm k,\bm p}\,,
\label{pfm-R}
\end{eqnarray}
Here, the Bunch-Davies vacuum $|0\rangle_{\rm I}$ and the vaccum in radiation-dominated era $|0\rangle_{\rm R}$ are defined respectively as 
\begin{eqnarray}
b^A_{\bm k}|0\rangle_{\rm I}=0\,,\qquad
c^A_{\bm k}|0\rangle_{\rm R}=0\,.
\label{vacua}
\end{eqnarray}
In the following, we omit the label of polarization modes $A$ and focus on either mode for simplicity because the equation of motion for different polarization modes are decoupled in the absence of the sources. 
From the relation between Eqs.~(\ref{pfm-I}) and (\ref{pfm-R}), we find the operators $b_{\bm k}$, $b^\dag_{-\bm k}$ and $c_{\bm k}$, $c^\dag_{-\bm k}$ are related by a Bogoliubov transformation
\begin{eqnarray}
b_{\bm k}=\alpha_k^{*} \,c_{\bm k} - \beta_k\,c_{-\bm k}^{\dag} \,,
\label{bogoliubov}
\end{eqnarray}
where the Bogoliubov coefficients can be read off from the relation as follows
\begin{eqnarray}
\alpha_k&=& \left(1-\frac{1}{2k^2\eta_1^2}-\frac{i}{k\eta_1}\right) e^{-2ik\eta_1}\quad\equiv\cosh r_k
\label{alpha}\,, \\
\beta_k&=& \frac{1}{2k^2\eta_1^2}\hspace{4.2cm}\equiv e^{i\varphi}\sinh r_k\,,
\label{beta}
\end{eqnarray}
so that $|\alpha_k|^2-|\beta_k|^2=1$ holds. The $\varphi$ is an arbitrary phase factor. Note that the Bogoliubov coefficients are written by a parameter $k\eta_1$. However, for later convenience, we introduced a new parameter $r_k$  known as the squeezing parameter. Applying Eq.~(\ref{bogoliubov}) to the definition of the Bunch-Davies vacuum in Eq.~(\ref{vacua}) and by using the commutation relations in Eq.~(\ref{pfm-I}), the Bunch-Davies vacuum can be written in terms of $c_{\bm k}$, $c^\dag_{\bm k}$ and the vacuum associated to each mode, $|0_{\bm k}\rangle_{\rm R}$ and $|0_{-\bm k}\rangle_{\rm R}$ such as
\begin{eqnarray}
|0\rangle_{\rm I}
= \prod_{\bm k}\sum_{n=0}^\infty e^{in\varphi}\frac{\tanh^nr_k}{\cosh r_k}\,
|n_{\bm k}\rangle_{\rm R}\otimes|n_{-\bm k}\rangle_{\rm R}\,,\qquad
\tanh r_k
=\biggl|\frac{\beta_k}{\alpha_k^{*}}\biggr|\,,
\label{two-mode1}
\end{eqnarray}
where we defined $|n_{\bm k}\rangle_{\rm R}=1/\sqrt{n!}\,(c_{\bm k}^{\dag})^n |0_{\bm k}\rangle_{\rm R}$ and 
$|0\rangle_{\rm R}=|0_{\bm k}\rangle_{\rm R}\otimes|0_{-\bm k}\rangle_{\rm R}$. The term $\cosh r_k$ is the normalization factor of this relation. In this way, $n$ particle excitation with momentum ${\bm k}$ and -${\bm k}$ appears. That is, the Bunch-Davies vacuum looks like graviton pair production occurs from the point of view of radiation-dominated era.

The rhs of Eq.~(\ref{two-mode1}) is obtained by applying the squeezing operator in Eq.~(\ref{squeezed}) to the vaccum of ${\bm k}$ and -${\bm k}$ modes in the radiation-dominated era as
\begin{eqnarray}
|0\rangle_{\rm I}
=\prod_{\bm k}\exp\left[\zeta^*c_{\bm k}\,c_{-\bm k}
-\zeta\,c_{\bm k}^{\dag}\,c_{-\bm k}^{\dag}\right]|0_{\bm k}\rangle_{\rm R}\otimes|0_{-\bm k}\rangle_{\rm R}=\prod_{\bm k}\hat{S} (\zeta)|0_{\bm k}\rangle_{\rm R}\otimes|0_{-\bm k}\rangle_{\rm R}\,,
\label{two-mode2}
\end{eqnarray}
where $\zeta=r_ke^{i\varphi}$. Thus, the Bunch-Davies vacuum is expressed by a two-mode squeezed state of the modes ${\bm k}$ and $-{\bm k}$ from the point of view of radiation-dominated era. Hence, we find that the Universe experienced a squeezed state in the past.

Furthermore, if we expand the exponential function in Eq.~(\ref{two-mode2}) in Taylor series, we find the two-mode squeezed state is an entangled state as below
\begin{eqnarray}
|0\rangle_{\rm I}\sim|0_{\bm k}\rangle_{\rm R}|0_{-\bm k}\rangle_{\rm R}+\zeta\,|1_{\bm k}\rangle_{\rm R}|1_{-\bm k}\rangle_{\rm R}+\zeta^2\,|2_{\bm k}\rangle_{\rm R}|2_{-\bm k}\rangle_{\rm R}\cdots\,.
\label{entangle}
\end{eqnarray}
That is, the Bunch-Davies vacuum looks like an entangled state between the modes ${\bm k}$ and $-{\bm k}$ of gravitons from the point of view of radiation-dominated era.

\subsection{The Bunch-Davies vacuum in the presence of matter fields}
\label{section4.2}

In the previous subsection, we found the Universe experienced the squeezed state in the past. In this subsection, we see how coherent state appears in the history of the Universe.

We consider the general action for the matter field. If we consider the linear interaction between metric and the matter field, we find it the definition of energy-momentum tensor $T^{\mu\nu}$ such as
\begin{eqnarray}
S_{\rm m}=\int d^4 x\,\frac{\delta S_{\rm m}}{\delta g_{\mu\nu} }\,\delta g_{\mu\nu}
=- \frac{1}{2} \int d^4 x  \sqrt{-g}\,T^{\mu\nu}\,\delta g_{\mu \nu}\,.
\end{eqnarray}
Then the interaction Hamiltonian becomes
\begin{eqnarray}
i \int d\eta H_{\rm int}&=&\frac{i}{2}\int d\eta\int d^3 x\,a^2(\eta) h_{ij}(\eta,{\bm x})\,T_{ij}(\eta,{\bm x})\,,\nonumber\\
&=&\sum_{\bm k}\sum_{A}
\left[\,
\xi_k^A\,b^{A\dag}_{{\bm k}}-\xi_k^{A*}\,b^A_{\bm k}
\,\right]\,,
\end{eqnarray}
where the coefficients $\xi_k^A$ is expressed as
\begin{eqnarray}
\xi_k^{A}=-\,\frac{i}{\sqrt{2}M_{\rm pl}} \int d\eta\,
a(\eta)\,p^{A}_{ij}({\bm k} )\,v_k^{{\rm I}*}(\eta)\,T_{ij} (\eta,-{\bm k} )\,.
\end{eqnarray}
Note that we used Eqs.~(\ref{fourier}) and (\ref{pfm-I}). This interaction generates a coherent state such as
\begin{eqnarray}
|\xi_k^A\rangle_{\rm I} &=& \exp\left[\,-i\int d\eta H_{\rm int}\,\right] |0\rangle_{\rm I}\,, \nonumber\\
 &=& \prod_{\bm k} \prod_{A}\exp\left[\,  
\xi_k^A\,b^{A\dag}_{{\bm k}}-\xi_k^{A*}\,b^{A}_{\bm k}      
\,\right] 
|0\rangle_{\rm I}
=\prod_{\bm k} \prod_{A}\hat{D}(\xi)|0\rangle_{\rm I}\,.
\end{eqnarray}
Hence, the Bunch-Davies vacuum in the presence of the matter fields becomes a coherent state~\cite{Glauber:1963fi}.

\subsection{Graviton statistics of the inflationary Universe}
\label{section4.3}

\begin{figure}[t]
\begin{center}
\vspace{-2cm}
\hspace{-2.1cm}
\begin{minipage}{8.0cm}
\includegraphics[height=7cm]{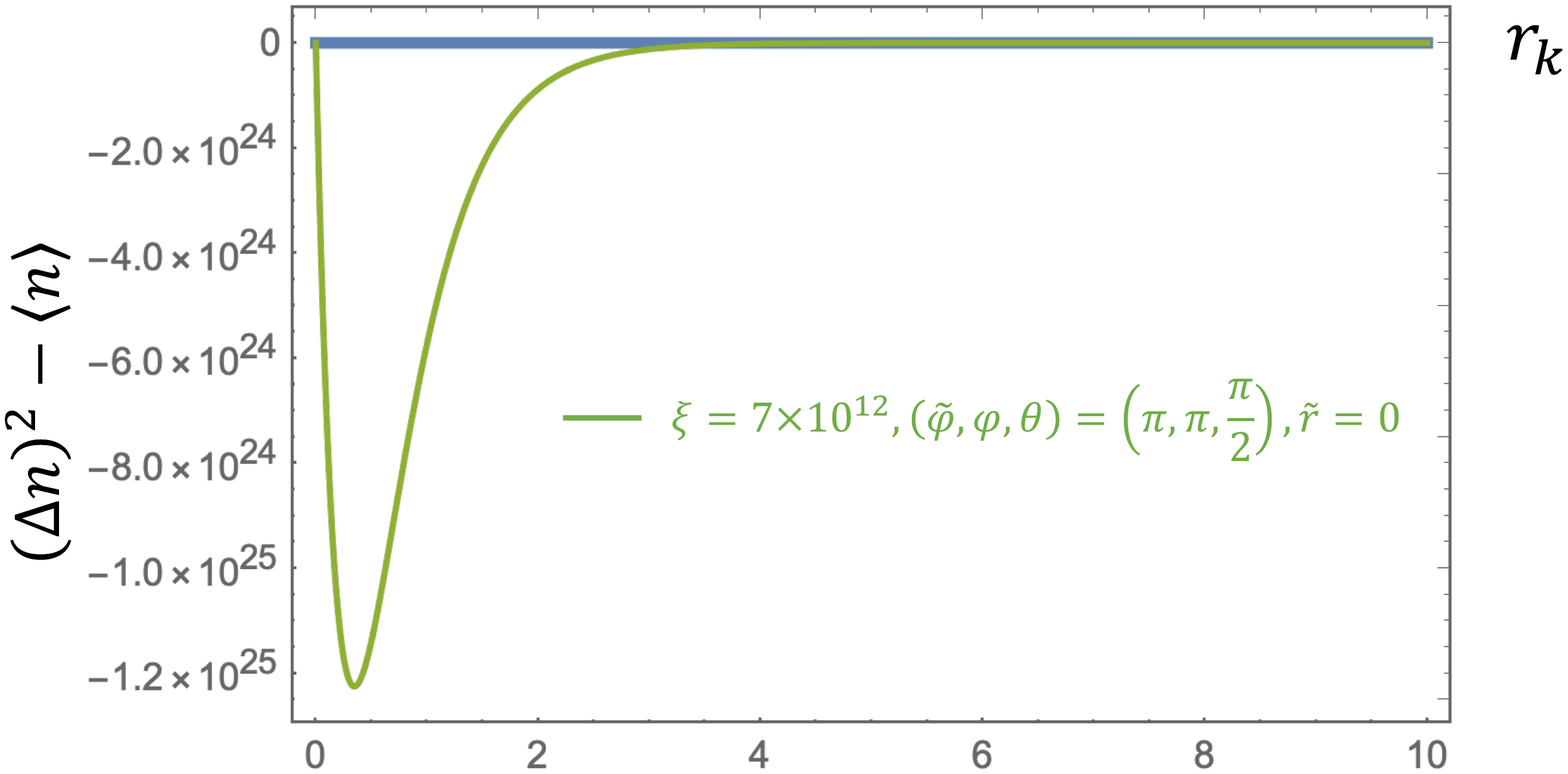}\centering
\end{minipage}
\begin{minipage}{8.0cm}
\hspace{0.7cm}
\includegraphics[height=7cm]{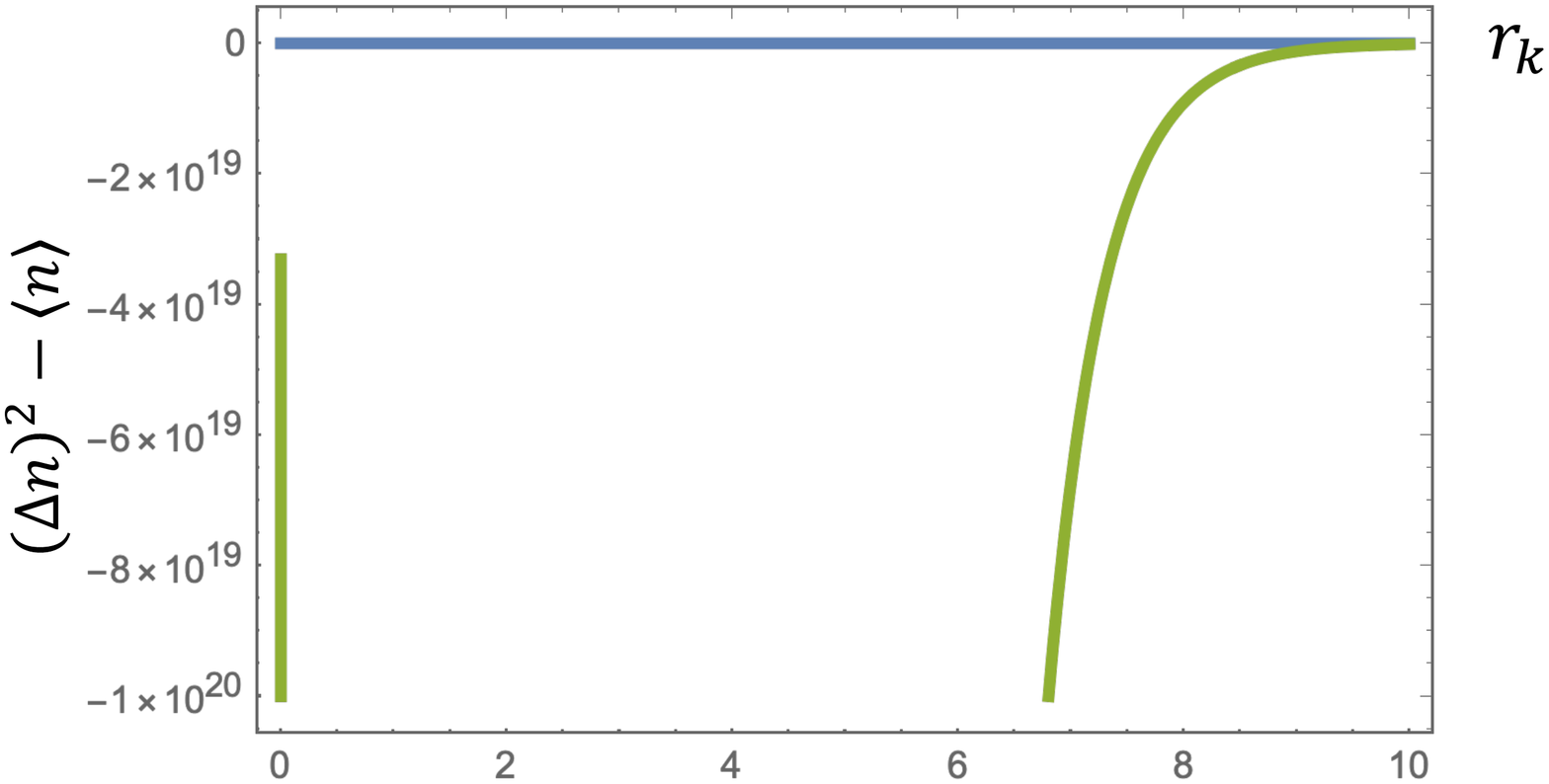}
\end{minipage}
\vspace{-0.8cm}
\caption{The plots of $\left(\Delta n\right)^2-\langle n\rangle$ versus squeezing parameter $r_k$ for the Bunch-Davies vacuum. The region $\left(\Delta n\right)^2-\langle n\rangle <0$  (below thick blue line) indicates that graviton statistics become sub-Poissonian. The enlarged plot around  $\left(\Delta n\right)^2-\langle n\rangle =0$ of the left panel is depicted in the right panel. The graviton statistics are always sub-Poissonian for $|\xi_k|\gg1$ if $\theta-\varphi/2=0$.}
\label{fig1}
\end{center}
\end{figure}

Up to here, we learnt that the Bunch-Davies vacuum looks like a squeezed state of gravitons from the point of view of radiation-dominated era. The initial presence of matter fields induce coherent state during inflation, which looks like a squeezed coherent state from the point of view of radiation-dominated era. In this subsection, we consider graviton statistics that an observer in radiation-dominated era finds. 

In the presence of matter fields, an observer in the vacuum state of radiation-dominated era will observe gravitons defined by operator $c_{\bm k}$. The expectation number of gravitons is found to be
\begin{eqnarray}
{}_{\rm I}\langle \xi_k|n_{\bm k}|\xi_k\rangle_{\rm I}
&=&{}_{\rm R}\langle \xi_k|\hat{S}^\dagger (\zeta)n_{\bm k}\hat{S} (\zeta)
|\xi_k\rangle_{\rm R}\,,\nonumber\\
&=&|\xi_k |^2\left[  e^{-2r_k} \cos^2\left(\theta -\frac{\varphi}{2} \right)
+e^{2r_k}\sin^2 \left(\theta -\frac{\varphi}{2}\right)
  \right] +\sinh^2r_k
\label{n}
\,,\\\nonumber
&=&{}_{\rm I}\langle \xi_k|n_{-\bm k}|\xi_k\rangle_{\rm I}\,,
\end{eqnarray}
where we used the fact that the coherent state in inflationary epoch looks like the squeezed coherent state from the point of view of radiation-dominated era in the first equality. We also used Eqs.~(\ref{coherent}), (\ref{relation1}) and (\ref{relation3}). The standard variance is
\begin{eqnarray}
\hspace{-5mm}
(\Delta n)^2&=&{}_{\rm I}\langle \xi_k| \left( n_{\bf k} +n_{- \bf k}\right)^2 
|\xi_k\rangle_{\rm I} 
-{}_{\rm I}\langle \xi_k|n_{\bm k}+n_{-\bm k}|\xi_k\rangle_{\rm I}^2\,,
\nonumber \\
&=&2|\xi_k |^2\left[e^{-4r_k}\cos^2\left(\theta -\frac{\varphi}{2} \right)
+e^{4r_k}\sin^2\left(\theta -\frac{\varphi}{2}\right)
\right]+4\sinh^2r_k+4\sinh^4r_k\,,
\end{eqnarray}
where we assumed that $n_{\bm k}$ and $n_{-\bm k}$ are indistinguishable and computed the standard variance for sum of them. Then the Fano factor Eq.~(\ref{fano}) becomes
\begin{eqnarray}
F &=&\frac{(\Delta n)^2}{{}_{\rm I}\langle \xi_k|\,n_{\bm k}+n_{-\bm k}\,|\xi_k\rangle_{\rm I}}\,,\nonumber\\
&=&\frac{|\xi_k |^2\left[e^{-4r_k}\cos^2\left(\theta-\frac{\varphi}{2}\right)
+e^{4r_k}\sin^2\left(\theta-\frac{\varphi}{2}\right)
\right]+2\sinh^2r_k+2\sinh^4r_k}
{|\xi_k |^2\left[e^{-2r_k}\cos^2\left(\theta-\frac{\varphi}{2}\right)
+e^{2r_k}\sin^2\left(\theta-\frac{\varphi}{2}\right)
\right]+\sinh^2r_k}\,.
\label{F}
\end{eqnarray}
For simplicity, if we take $\theta-\varphi/2=0$, we get
\begin{eqnarray}
F =\frac{|\xi_k |^2e^{-4r_k}+ 2\sinh^2r_k+2\sinh^4r_k}
{|\xi_k |^2e^{-2r_k}+\sinh^2r_k}\,.
\label{fano2}
\end{eqnarray}
If the Fano factor satisfies
\begin{eqnarray}
|\xi_k |^2 \left(e^{-2r_k} - e^{-4r_k}\right) >\sinh^2 r_k+2\sinh^4 r_k\,,
\label{condition}
\end{eqnarray}
the graviton statistics in the squeezed coherent state become sub-Poissonian. Note that we have $F\sim e^{-2r}<1$ for $|\xi_k|\gg1$ in Eq.~(\ref{fano2}). Thus, the graviton statistics are always sub-Poissonian for $|\xi_k|\gg1$ if $\theta-\varphi/2=0$. This is plotted in Figure~\ref{fig1}. 

\subsection{Frequency range of nonclassical PGWs}
\label{section4.4}

In this subsection, we rewrite the condition Eq.~(\ref{condition}) in terms of frequency range of nonclassical PGWs. Since the relation between $k\eta_1$ and $r_k$ is given in Eq.~(\ref{beta}), we first focus on $k\eta_1$. We translate the comoving wave number $k$ into physical wave number and the time inflation ends $\eta_1$ into physical frequency at present. Then the quantity $k|\eta_1|$ is computed as
\begin{eqnarray}
k|\eta_1 | \equiv  \frac{f}{f_1}\,,\qquad
f_1 = 10^9 \sqrt{\frac{H}{10^{-4}M_{\rm pl}}}\quad[{\rm Hz}]\,,
\end{eqnarray}
where $f_1$ is the cutoff frequency. Thus, there are no more PGWs generated during inflation with frequency higher than 1 GHz when the Hubble parameter is $10^{-4} M_{\rm pl}$. Pluggin this back into Eq.~(\ref{beta}), we have
\begin{eqnarray}
\sinh r_k=\frac{1}{2}\left(\frac{f_1}{f}\right)^2\,.
\end{eqnarray}
Combining the above relation with the condition Eq.~(\ref{condition}), we get the condition to observe nonclassical PGWs can be approximately written by
\begin{eqnarray}
f  > \left(\frac{1}{8}\right)^{\frac{1}{12}}10^9\,|\xi_k |^{-\frac{1}{6}}
\sqrt{\frac{H}{10^{-4}M_{\rm pl}}}\quad[{\rm Hz}] \,.
\label{frequency}
\end{eqnarray}
Since 1 GHz is a cutoff scale for PGWs generated during inflation, we have the chance to observe the nonclassical PGWs if the amplitude of $|\xi_k|$ is larger than 1.

\subsection{Prediction of frequency range of nonclassical PGWs}
\label{section4.5}

In our previous paper~\cite{Kanno:2018cuk}, the $|\xi_k|$ of Eq.~(\ref{frequency}) was estimated by considering two models with a gauge field as the matter field during inflation (Anisotropic inflation model~\cite{Watanabe:2009ct, Soda:2012zm, Maleknejad:2012fw} and Axion inflation model~\cite{Barnaby:2010vf,Barnaby:2011vw,Cook:2011hg,Anber:2012du,Barnaby:2011qe}). In both models, gauge fields grow during inflation and disappear after the inflation. In the anisotropic inflation model~\cite{Choi:2015wva, Ito:2016aai}l, the frequency range in which we can observe nonclassicality is given by
\begin{eqnarray}
f>10^{8.1}\,e^{-\frac{4}{17}\nu N_{\rm gauge}}
\left( \frac{H}{10^{-4}M_{\rm pl}}\right)^{\frac{6}{17}}   \,{\rm [Hz]}\,.
\end{eqnarray}
Then the nonclassical PGWs can be observed for $f > 100 \  {\rm kHz}$
with the model parameter $\nu N_{\rm gauge} \sim 30$ and $H= 10^{-4}M_{\rm pl}$. On the other hand, in the axion inflation model, the frequency range is found to be
\begin{eqnarray}
 f>  {10^{7.9}}\,e^{-\frac{2}{7}\pi\chi }\chi^{\frac{1}{14}} \left( \frac{H}{10^{-4}M_{\rm pl}}\right)^{\frac{9}{28}}   \,{[\rm Hz]}\,.
\end{eqnarray}
If we take the model parameter $\chi\sim 10$ and $H= 10^{-4}M_{\rm pl}$, the frequency range reduces to $f>10$ kHz, which can be marginally observed nonclassicality in the PGWs with the LIGO detector.

\subsection{A remark on necessary condition for phases}
\label{section4.6}

In Eq.~(\ref{fano2}), we considered the case of $\theta-\varphi/2=0$ for simplicity. However this combination of phases is another necessary condition to have a chance to get sub-Poissonian graviton statistics. This is because, in the exact form of Fano factor in Eq.~(\ref{F}), the second term in the numerator of $F$ becomes dominant for $|\xi_k|\gg1$ and $r_k\gg1$ if $\theta-\varphi/2\neq0$. Then we have $F>1$ and graviton statistics become super-Poissonian. The $\theta$ can be zero in the two models of anisotropic and axion inflation models. However, we cannot set $\varphi=0$ in Eq.~(\ref{alpha}) because $\eta_1$ has to be finite values, so as not to get the squeezing limit $\eta_1\rightarrow 0$ ($r_k\rightarrow\infty$) which corresponds to super-Poissonian statistics. Thus we need the combination of the phases $\theta-\varphi/2=0$ in order to observe nonclassical PGWs if the initial state of the Universe is the Bunch-Davies vacuum\footnote{If $\theta-\varphi/2\neq0$, the result would agree with the papers~\cite{Giovannini:2019bfw, Giovannini:2019ehc}}.

\section{Graviton statistics in the initial entangled state}
\label{section5}

In the previous section, we reviewed the graviton statistics in the Bunch-Davies vacuum. Here, we extend the initial state to more general de Sitter invariant vacua, that is, $\alpha$-vacua. The $\alpha$-vacua look like entangled states from the point of view of the Bunch-Davies vacuum as shown in the following.

\subsection{Initial entagled states -- $\alpha$-vacua}
\label{section5.1}

Suppose that the Universe starts from $\alpha$-vacua, then the operator $h_{\bm k}(\eta)$ is expanded as
\begin{eqnarray}
h_{\bm k}=d_{\bm k}\,v^{\rm E}_k(\eta)+d_{-\bm k}^\dag\,v_k^{\rm E *}(\eta)\,,\qquad
\left[d_{\bm k} , d_{\bm p}^\dag \right]= \delta_{\bm k,\bm p}\,,
\label{pfm-E}
\end{eqnarray}
where $v^{\rm E}_k(\eta)$ is the positive frequency mode in the $\alpha$-vacua which is obtained by the Bogoliubov transformation from the positive frequency mode of the Bunch-Davies vacuum $v_k^{\rm I}$ in Eq.~(\ref{modefunction}) sucn as
\begin{eqnarray}
v^{\rm E}_k(\eta)=\cosh\tilde{r}_k\,v^{\rm I}_k(\eta)+\sinh\tilde{r}_k\,v^{\rm I*}_k(\eta)\,.
\label{pfm-E}
\end{eqnarray}
Here, $\tilde{r}_k$ is the squeezing parameter in the $\alpha$-vacua.
Comparing Eq.~(\ref{pfm-E}) with Eq.~(\ref{pfm-I}), we find the operators $d_{\bm k}$, $d_{-\bm k}$ and $b_{\bm k}$, $b_{-\bm k}$ are related by a Bogoliubov transformation
\begin{eqnarray}
d_{\bm k}=\gamma_k^{*} \,b_{\bm k} - \delta_k\,b_{-\bm k}^{\dag}\,,
\end{eqnarray}
where the Bogoliubov coefficients can be written as
\begin{eqnarray}
\gamma_k=\cosh\,\tilde{r}_k\,,\qquad\delta_k=e^{i\tilde{\varphi}}\sinh\,\tilde{r}\,.
\end{eqnarray}
Here, $\tilde{\varphi}$ is an arbitrary phase factor. Then the $\alpha$-vacua are expressed in terms of the Bunch-Davies vacuum as
\begin{eqnarray}
|0\rangle_{\rm E}
&=& \prod_{\bm k}\sum_{n=0}^\infty e^{in\tilde{\varphi}}\frac{\tanh^n\tilde{r}_k}{\cosh\tilde{r}_k}\,
|n_{\bm k}\rangle_{\rm I}\otimes|n_{-\bm k}\rangle_{\rm I}\,,\qquad
\tanh\tilde{r}_k
=\biggl|\frac{\delta_k}{\gamma_k^{*}}\biggr|\,,\nonumber\\
&=&\prod_{\bm k}\exp\left[\tilde{\zeta}^*b_{\bm k}\,b_{-\bm k}
-\tilde{\zeta}\,b_{\bm k}^{\dag}\,b_{-\bm k}^{\dag}\right]|0_{\bm k}\rangle_{\rm I}\otimes|0_{-\bm k}\rangle_{\rm I}=\prod_{\bm k}\hat{U} (\tilde{\zeta})|0_{\bm k}\rangle_{\rm I}\otimes|0_{-\bm k}\rangle_{\rm I}\,,
\end{eqnarray}
where $\tilde{\zeta}=\tilde{r}_ke^{i\tilde{\varphi}}$. $\hat{U}$ is the squeezing operator. As we reviewed in Section~\ref{section4.1}, this can be expanded in the form of an entangled state as in  Eq.~(\ref{entangle}). Thus, the $\alpha$-vacua look like entangled states from the point of view of the Bunch-Davies vacuum.

\subsection{Graviton statistics}
\label{section5.2}

In subsection~\ref{section4.3} , we found the necessary condition for graviton statistics to become sub-Poissonian in the case of the Bunch-Davies vacuum. In this subsection, we assume the presence of matter fields in the initial entangled state and consider 
graviton statistics that an observer in radiation-dominated era finds.

The expectation number of gravitons is calculated as
\begin{eqnarray}
{}_{\rm E}\langle \xi_k|n_{\bm k}|\xi_k\rangle_{\rm E}
&=&{}_{\rm I}\langle \xi_k|\hat{U}^\dagger (\tilde{\zeta})n_{\bm k}\hat{U} (\tilde{\zeta})
|\xi_k\rangle_{\rm I}
={}_{\rm R}\langle \xi_k|\hat{S}^\dagger (\zeta)\hat{U}^\dagger (\tilde{\zeta})n_{\bm k}\hat{U} (\tilde{\zeta})\hat{S} (\zeta)
|\xi_k\rangle_{\rm R}\nonumber\\
&=&|\xi_k |^2\Bigl(\,
|f|^2+|g|^2-e^{i\left(\varphi-2\theta\right)}f^*g-e^{-i\left(\varphi-2\theta\right)}fg^*
\,\Bigr) +|g|^2\nonumber\\
&=&{}_{\rm E}\langle \xi_k|n_{-\bm k}|\xi_k\rangle_{\rm E} \,,
\end{eqnarray}
where we defined
\begin{eqnarray}
f&\equiv&\Bigl(\cosh\tilde{r}_k+\sin\left(\tilde{\varphi}-\varphi\right)\sinh\tilde{r}_k\sinh2r_k\Bigr)\cosh r_k\nonumber\\
&&\qquad+\sinh\tilde{r}_k\Bigl(e^{i\left(\tilde{\varphi}-\varphi\right)}\cosh^2r_k+e^{-i\left(\tilde{\varphi}-\varphi\right)}\sinh^2r_k\Bigr)\sinh r_k\,,\nonumber\\
g&\equiv&\Bigl(\cosh\tilde{r}_k+\sin\left(\tilde{\varphi}-\varphi\right)\sinh\tilde{r}_k\sinh2r_k\Bigr)\sinh r\nonumber\\
&&\qquad+\sinh\tilde{r}_k\Bigl(e^{i\left(\tilde{\varphi}-\varphi\right)}\cosh^2r_k+e^{-i\left(\tilde{\varphi}-\varphi\right)}\sinh^2r_k\Bigr)\cosh r_k\,.
\end{eqnarray}
Note that $\tilde{\varphi}=\varphi$ and $\tilde{r}_k=0$ correspond to the case of the Bunch-Davies vacuum and then we have $f=\cosh r_k$ and $g=\sinh r_k$ which recover  Eq.~(\ref{n}). Here, we used the fact that the coherent state in the initial entangled state looks like the squeezed coherent state from the point of view of the Bunch-Davies vacuum in the first equality. Then we used Eqs.~(\ref{relation1}), (\ref{relation2}) and (\ref{relation4}). The standard variance in this case is calculated as
\begin{eqnarray}
(\Delta n)^2&=&{}_{\rm E}\langle \xi_k| \left( n_{\bf k} +n_{- \bf k}\right)^2 
|\xi_k\rangle_{\rm E} 
-{}_{\rm E}\langle \xi_k|n_{\bm k}+n_{-\bm k}|\xi_k\rangle_{\rm E}^2\nonumber\\
&=&|\xi_k|^2\Bigl(6|fg|^2+2|g|^4+|f|^2+|g|^2-\left(|f|^2+2|g|^2+1\right)\left(e^{i\left(\varphi-2\theta\right)}f^*g+e^{-i\left(\varphi-2\theta\right)}fg^*\right)\Bigr)\nonumber\\
&&\quad+|fg|^2+|g|^4+|g|^2\,.
\end{eqnarray}

Finally, we find the difference between the standard variance and the expectation number is expressed as
\begin{eqnarray}
\left(\Delta n\right)^2-{}_{\rm E}\langle \xi_k|n_{\bm k}+n_{-\bm k}|\xi_k\rangle_{\rm E}
=|\xi_k |^2A+B\,,
\label{comb}
\end{eqnarray} 
where $A$ and $B$ consist of
\begin{eqnarray}
A&=&AO16+AO14+AO12+AO10+AO8\,,\nonumber\\
B&=&BO16+BO14+BO12+BO10+BO8\,.
\end{eqnarray}
Here $O16$ and so forth represent the total order of squeezing parameters. The details of  $AO16,\cdots ,AO8$ and $BO16\cdots ,BO8$ are given in Appendix~\ref{appC}.

\subsection{Nonclassical PGWs from the initial entangled state}
\label{section5.3}

\begin{figure}[t]
\begin{center}
\vspace{-2cm}
\hspace{-2.1cm}
\begin{minipage}{8.0cm}
\includegraphics[height=7cm]{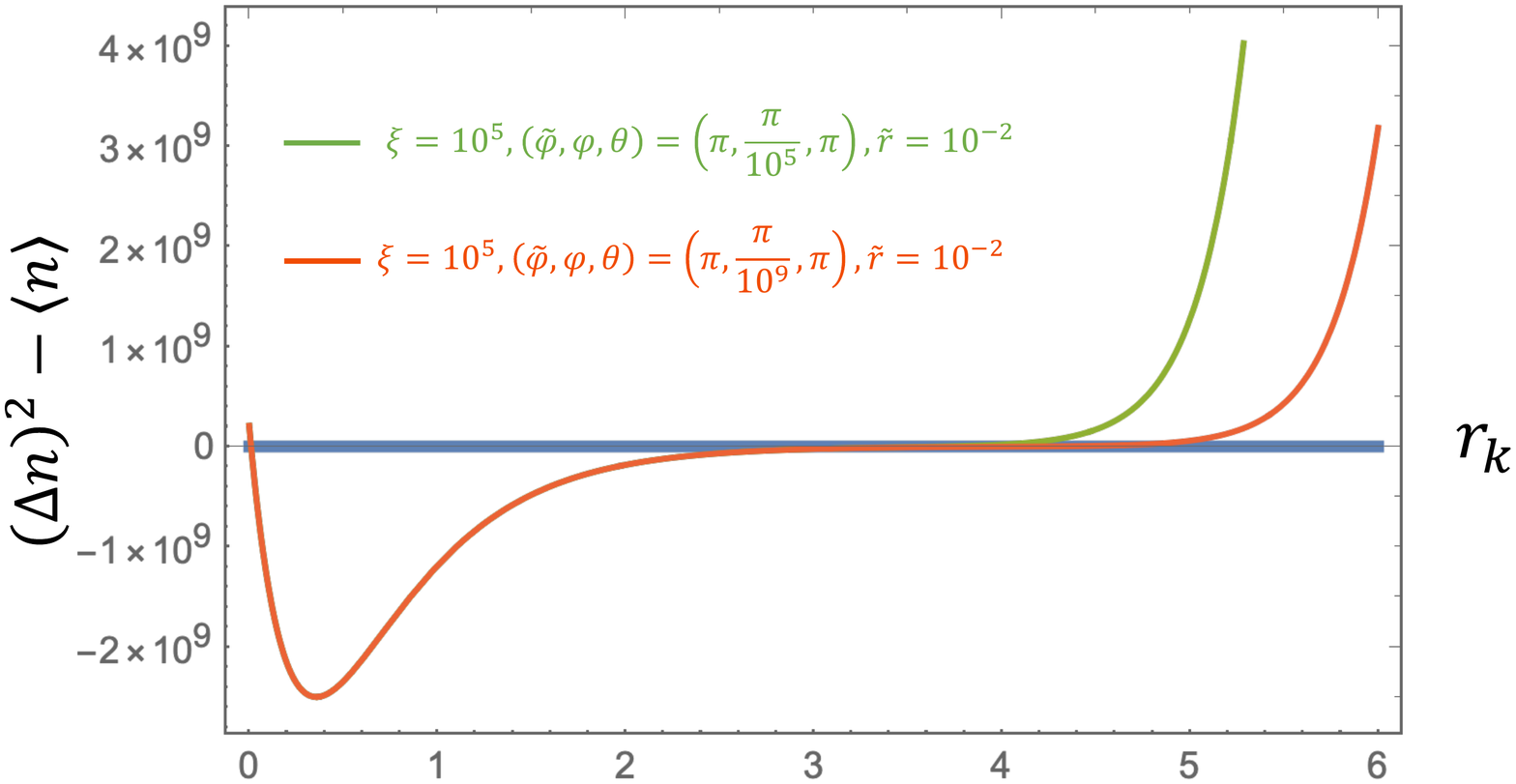}\centering
\end{minipage}
\begin{minipage}{8.0cm}
\hspace{0.7cm}
\includegraphics[height=7cm]{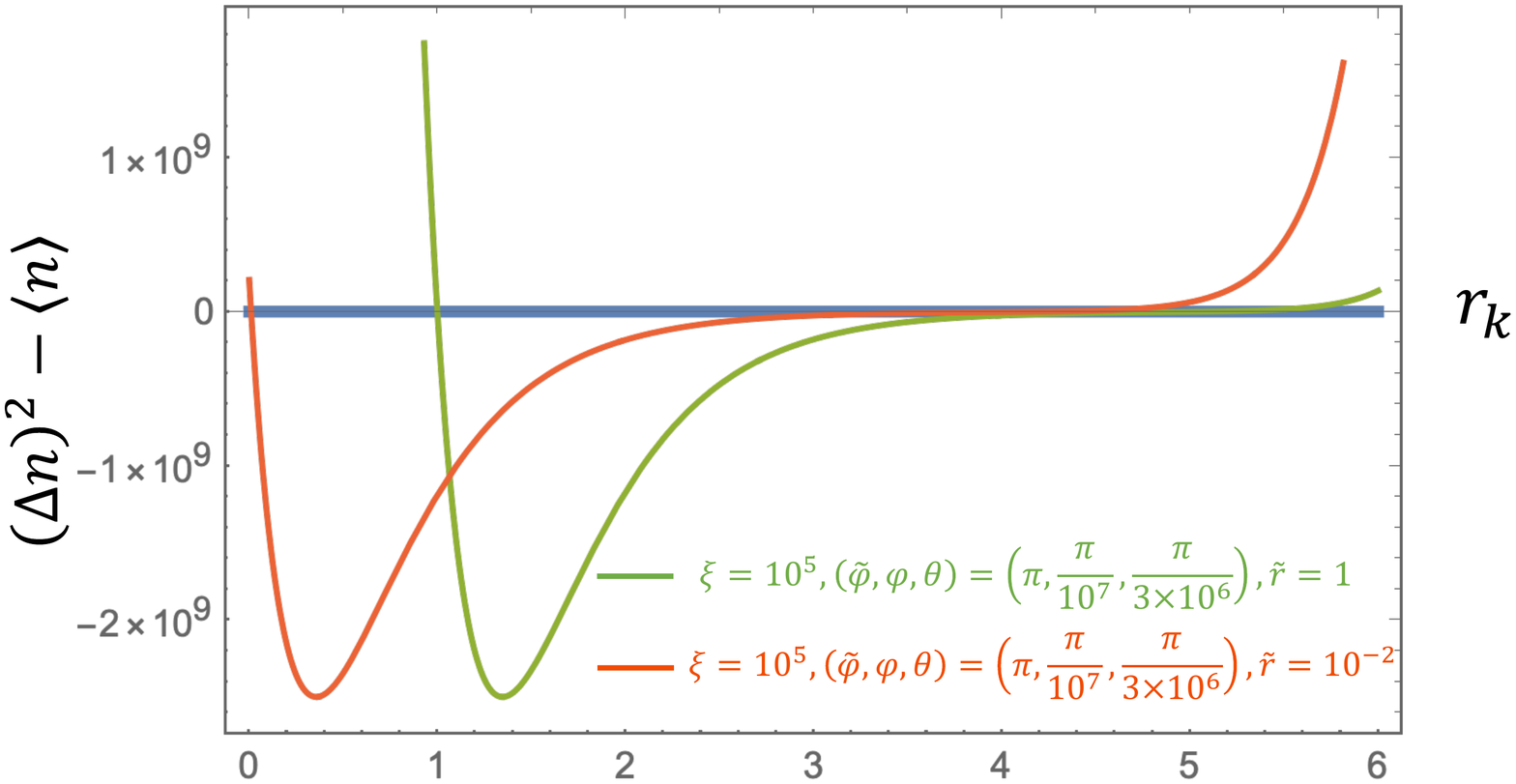}
\end{minipage}
\vspace{-0.8cm}
\caption{The plots of $\left(\Delta n\right)^2-\langle n\rangle$ versus squeezing parameter $r_k$ for the initial entagled state. The region $\left(\Delta n\right)^2-\langle n\rangle <0$  (below thick blue line) indicates that graviton statistics become sub-Poissonian. The left panel shows the sub-Poisson range increases as the difference between $\tilde\varphi$ and $\varphi$ gets smaller. The difference is $\pi\left(1-1/10^5\right)$ (green) and $\pi\left(1-1/10^9\right)$ (red).  The right panel shows the sub-Poissonian range shifts to the large value of $r_k$ keeping the same shape as we increase $\tilde{r}_k$.}
\label{fig2}
\end{center}
\end{figure}

As we discussed in Eq.~(\ref{condition}), the condition for graviton statistics to become sub-Poissonian is that the Fano factor in Eq.~(\ref{fano}) satisfies $F<1$, that is, the rhs of Eq.~(\ref{comb}) is
\begin{eqnarray}
|\xi|^2A+B<0\,.
\label{graviton}
\end{eqnarray}
Note that the above condition recovers Eq.~(\ref{F}) for $\tilde{\varphi}-\varphi=0$ and $\tilde{r}=0$. The condition for the graviton statistics to become sub-Poissonian is depicted in Figure~\ref{fig2}. Below the thick blue line $\left(\Delta n\right)^2-\langle n\rangle=|\xi|^2A+B<0$ indicates that graviton statistics is sub-Poissonian. The range of squeezing parameter during inflation $r_k$ turns out to be frequency range for the nonclassical PGWs as in Eq.~(\ref{frequency}). 
The left panel shows the dependence of $\tilde{\varphi}-\varphi$ on $\left(\Delta n\right)^2-\langle n\rangle$. We see that squeezing parameter $r_k$ tends to increase as $\tilde{\varphi}-\varphi$ gets smaller. On the right panel, we depicted the dependence of the squeezing parameter of the initial entangled state $\tilde{r}$ on $\left(\Delta n\right)^2-\langle n\rangle$. As $\tilde{r}_k$ increases, the sub-Poissonian range tends to shift to larger value of $r_k$ keeping its shape.

\begin{figure}[t]
\begin{center}
\vspace{-2cm}
\hspace{-2.1cm}
\begin{minipage}{8.0cm}
\includegraphics[height=7cm]{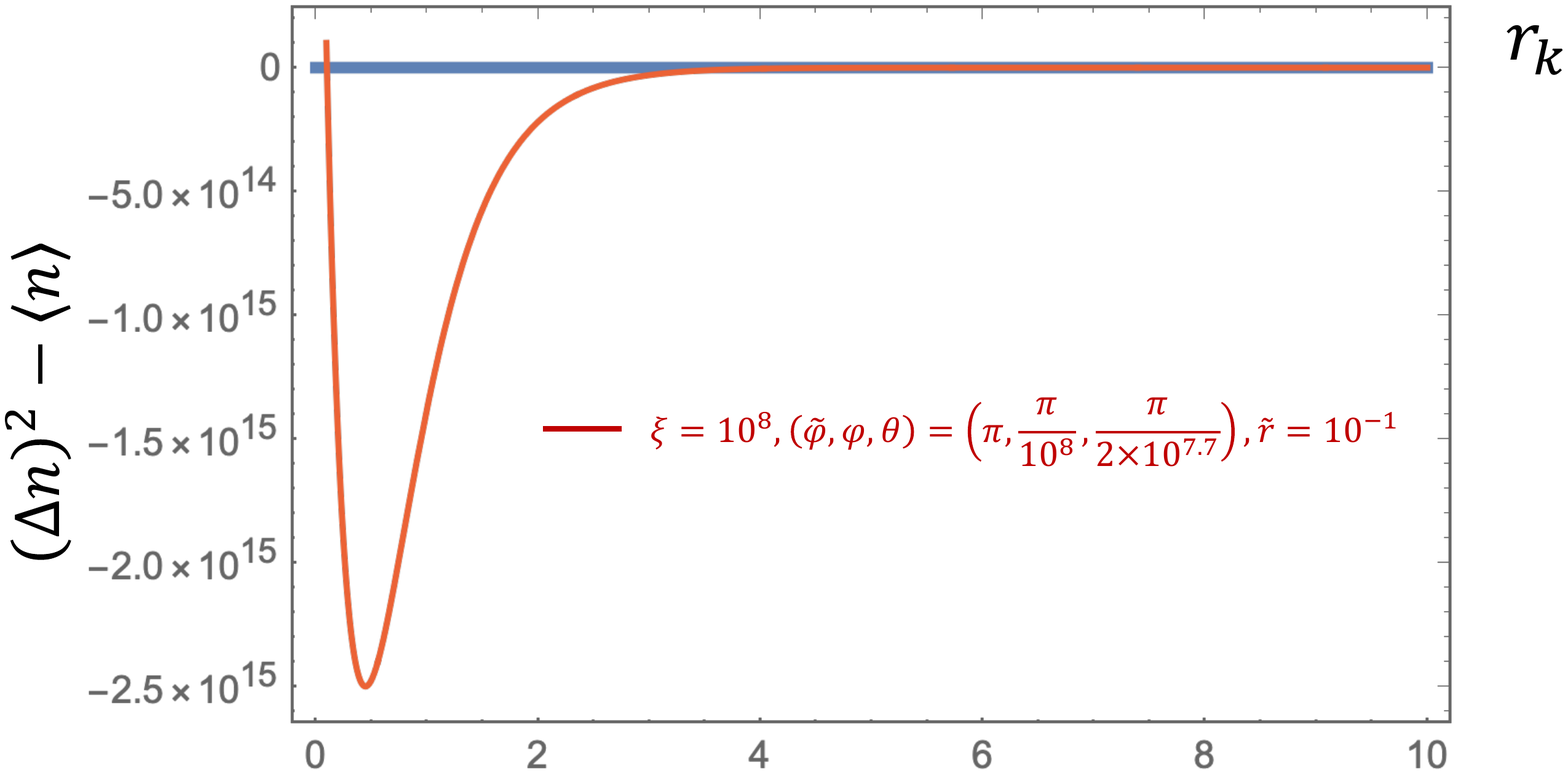}\centering
\end{minipage}
\begin{minipage}{8.0cm}
\hspace{0.7cm}
\includegraphics[height=7cm]{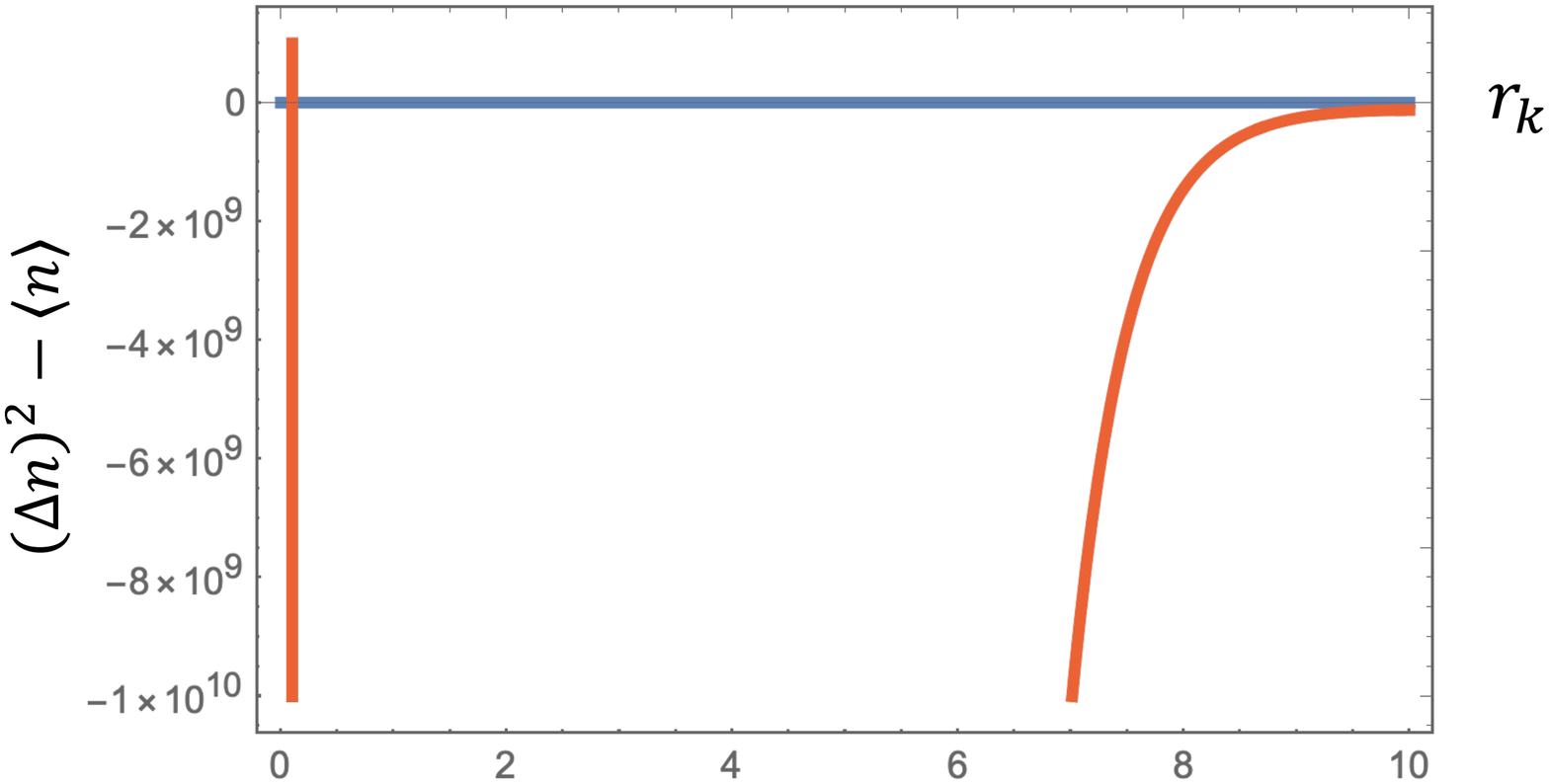}
\end{minipage}
\vspace{-0.8cm}
\caption{The plots of $\left(\Delta n\right)^2-\langle n\rangle$ versus squeezing parameter $r_k$ for the Bunch-Davies vacuum. The region $\left(\Delta n\right)^2-\langle n\rangle <0$  (below thick blue line) indicates that graviton statistics become sub-Poissonian. The right panel is enlarged plot of the left panel around  $\left(\Delta n\right)^2-\langle n\rangle =0$. The graviton statistics become sub-Poissonian for $\tilde{\varphi}\sim\varphi$ with smaller values of $|\xi_k|$ than the values of Bunch-Davies vacuum and $\theta\neq\phi/2$.}
\label{fig3}
\end{center}
\end{figure}

Let us discuss the condition for graviton statistics to be sub-Poissonian in Eq.~(\ref{graviton}). If we take large coherence $|\xi_k |\gg 1$ and squeezing $r_k\gg 1$ during inflation, then the $|\xi_k |^2AO16$ term becomes dominant. The graviton statistics then become super-Poissonian because $AO16$ in Eq.~(\ref{ao16}) is positive definite. However, we can think of the situation where the squeezing is not so strong and just $r_k>1$. Then the next order $AO14$ overcomes the $AO16$ in some cases. In such a situation, the graviton statistics can be sub-Poissonian. This situation occurs if the squeezing parameter satisfies the condition
\begin{eqnarray}
|\xi_k |^2\left(AO16+AO14\right)+BO16<0\,,
\label{condition2}
\end{eqnarray}
which is written as
\begin{eqnarray}
e^{2r_k}<\frac{8\,|\xi_k |^2\sin\left(\frac{\varphi}{2}-\theta\right)\Bigl(\,3\sin\left(\theta-\frac{\tilde\varphi}{2}\right)+\sin\left(\theta+\frac{\tilde\varphi}{2}-\varphi\right)\Bigl)}{\sin\left(\frac{\tilde{\varphi}}{2}-\frac{\varphi}{2}\right)\Bigl(\,8|\xi_k|^2\sin^2\left(\frac{\varphi}{2}-\theta\right)+1\,\Bigr)}\,,
\label{condition3}
\end{eqnarray}
where we approximated Eq.~(\ref{condition2}) by taking large enough $r_k$. The squeezing parameter in the entangled state $\tilde{r}_k$ is canceled out.
The above condition tells us that unlike the condition of Bunch-Davies vacuum in Eq.~(\ref{F}), we have a chance to have sub-Poissonian statistics even if $\theta\neq\varphi/2$. And if $\tilde{\varphi}\sim\varphi$, then the sub-Poissonian range of $r_k$ increases as long as Eq.~(\ref{condition3}) is satisfied as is shown in the left panel of Figure~\ref{fig2}. This range of squeezing parameter $r_k$ turns out to be frequency range for the nonclassical PGWs as in Eq.~(\ref{frequency}). For $|\xi|\gg 1$, Eq.~(\ref{condition3}) become
\begin{eqnarray}
e^{2r_k}<\frac{3\sin\left(\theta-\frac{\tilde\varphi}{2}\right)+\sin\left(\theta+\frac{\tilde\varphi}{2}-\varphi\right)}{\sin\left(\frac{\tilde{\varphi}}{2}-\frac{\varphi}{2}\right)\sin\left(\frac{\varphi}{2}-\theta\right)}\,.
\label{condition4}
\end{eqnarray}
Hence, unlike the condition mentioned below of Eq.~(\ref{condition}), the above condition does not depend on $|\xi_k|$. If we take $\tilde{\varphi}\sim\varphi$ or $\theta\sim\varphi/2$, the sub-Poissonian range of $r_k$ increases as long as Eq.~(\ref{condition4}) is satisfied. This is depicted in Figure~\ref{fig3}.

\section{Summary and discussion}
\label{section6}

We explored the conditions for primordial gravitational waves (PGWs) to be nonclassical. We characterized the nonclassicality by sub-Poissonian graviton statistics. Among quantum states, we find that squeezed coherent states realize the sub-Poissonian statistics. In our previous work~\cite{Kanno:2018cuk}, we studied the graviton statistics when the initial quantum state is the Bunch-Davies vacuum. We found that the presence of matter fields during inflation makes graviton statistics sub-Poissonian. We derived the condition for graviton statistics to be sub-Poissonian in Eq.~(\ref{condition}), which tells us that the modes of PGWs that do not stay long time outside horizon tend to be sub-Poissonian. We concluded that PGWs with frequency higher than $10$ kHz enable us to observe their nonclassicality. Besides the condition, another condition between the phases of squeezing and coherent parameters was necessary to have sub-Poissonian graviton statistics. 

In this work, we extended the initial quantum state to entangled states. As the initial entangled state, we considered $\alpha$-vacua, which are more general de Sitter invariant vacua than the Bunch-Davies vacuum. We found that, unlike the Bunch-Davies vacuum, the nonclassical PGWs generated in the initial entangled state become sub-Poissonian withouth requiring the condition between the phases and tend to keep their nonclassicaity outside the horizon as long as Eq.~(\ref{condition3}) is satisfied.

Let us discuss the possible detection of the nonclassical PGWs. In quantum optics, it is known that the sub-Poissonian statistics can be detected with Hanbury Brown and Twiss (HBT) interferometry~\cite{HanburyBrown:1956bqd, Brown:1956zza}. The HBT interferometry is a method to investigate the nonclassical nature of fields developed in quantum optics. This concept has been first applied to cosmology in~\cite{Giovannini:2010xg,Giovannini:2016esa,Giovannini:2017uty} and more recently in~\cite{Chen:2017cgw}. The HBT interferometry considers two-source interference and uses two detectors. The signals from the two detectors are converted to electronically correlated current and the net current is measured. The HBT interferometry measures the intensity-intensity correlations characterized by the second-order coherent function $g^{(2)} $
\begin{eqnarray}
g^{(2)} (\tau)= \frac{\langle a^\dagger (t) a^\dagger (t+\tau) a (t+\tau)a (t)\rangle}{\langle a^\dagger (t)a (t)\rangle \langle a^\dagger (t+\tau)a (t+\tau)\rangle}\,,
\end{eqnarray}
where the time delay between the two signals at the two detectors is expressed by $\tau$. If the sources are classical, the operators $a$ and $a^\dag$ become the amplitude of the two fields. This second order coherence function makes us possible to distinguish between classical and nonclassical fields from the fringe pattern of the interference. The point is that the second order coherence function can be expressed by using the Fano factor in Eq.~(\ref{fano}) as follows
\begin{eqnarray}
g^{(2)} (0) =1+\frac{\left( \Delta n \right)^2-\langle n \rangle }{\langle n   \rangle^2}
=1+ \frac{F - 1 }{\langle n \rangle }\,.
\end{eqnarray}
Hence, if the sources are classical fields, the Fano factor is above one and then $g^{(2)}$ becomes larger than one. On the other hand, if the sources are nonclassical fields, the Fano factor is below one and then $g^{(2)}$ becomes smaller than one. In this way, we can distinguish between classical and nonclassical fields by using the HBT interferometry. Thus, we could detect the nonclassical PGWs with the HBT interferometry if an experiment is carried out to detect nonclassical PGWs in the future.

\section*{Acknowledgments}

I would like to thank Jiro Soda and Jon Shock for useful discussions. This work was supported by JSPS KAKENHI Grant No. JP19K03827.

\appendix

\section{Some formulas}
\label{appA}

If we use the relation,
$e^AB\,e^{-A}=B+[A,B]+1/(2!)[A,[A,B]]+\cdots$, we find Eq.~(\ref{squeezed}) leads to
\begin{eqnarray}
{\hat S}^\dag(\zeta_k)\,c_{\bm k}\,{\hat S}(\zeta_k)&=&c_{\bm k}\cosh r_k-c_{-\bm k}^\dag\,e^{i\varphi}\sinh r_k\,,
\nonumber\\
{\hat S}^\dag(\zeta_k)c^\dag_{\bm k}\,{\hat S}(\zeta_k)&=&c^\dag_{\bm k}\cosh r_k-c_{-\bm k}
\,e^{-i\varphi}\sinh r_k\,.
\label{relation1}
\end{eqnarray}
Similarly, we have 
\begin{eqnarray}
{\hat U}^\dag(\tilde{\zeta_k})\,b_{\bm k}\,{\hat U}(\tilde{\zeta_k})&=&b_{\bm k}\cosh\tilde{r}_k-b_{-\bm k}^\dag\,e^{i\tilde{\varphi}}\sinh\tilde{r}_k\,,
\nonumber\\
{\hat U}^\dag(\tilde{\zeta_k})b^\dag_{\bm k}\,{\hat U}(\tilde{\zeta_k})&=&b^\dag_{\bm k}\cosh\tilde{r}_k-b_{-\bm k}
\,e^{-i\tilde{\varphi}}\sinh\tilde{r}_k\,.
\end{eqnarray}
By using the Bogoliubov transformation $b_{\bm k}=\alpha^*c_{\bm k}-\beta c^\dag_{-\bm k}$, we get
\begin{eqnarray}
{\hat U}^\dag(\tilde{\zeta_k})\,c_{\bm k}\,{\hat U}(\tilde{\zeta_k})&=&c\,\Bigl(\cosh\tilde{r}_k+\sin\left(\tilde{\varphi}-\varphi\right)\sinh2r_k\sinh\tilde{r}_k\Bigr)\nonumber\\
&&\qquad-c^\dag e^{i\varphi}\sinh\tilde{r}_k\Bigl(e^{i\left(\tilde{\varphi}-\varphi\right)}\cosh^2r_k-e^{-i\left(\tilde{\varphi}-\varphi\right)}\sinh^2r_k\Bigr)\,,\nonumber\\
{\hat U}^\dag(\tilde{\zeta_k})\,c^\dag_{\bm k}\,{\hat U}(\tilde{\zeta_k})&=&c^\dag\Bigl(\cosh\tilde{r}_k+\sin\left(\tilde{\varphi}-\varphi\right)\sinh2r_k\sinh\tilde{r}_k\Bigr)\nonumber\\
&&\qquad-c\,e^{-i\varphi}\sinh\tilde{r}_k\Bigl(e^{-i\left(\tilde{\varphi}-\varphi\right)}\cosh^2r_k-e^{i\left(\tilde{\varphi}-\varphi\right)}\sinh^2r_k\Bigr)\,.
\label{relation2}
\end{eqnarray}

\section{Relation between coherent and squeezing operators}
\label{appB}

The displacement operator has  a relation below
\begin{eqnarray}
\hat{D}^{\rm I}\left(\xi_k\right)=\exp\left[\xi_k b^\dag-\xi^*b_k\right]=\exp\left[\bar{\xi}_kc^\dag-{\bar\xi}_k^*c\right]=\hat{D}^{\rm R}\left(\bar{\xi}_k\right)\,,
\end{eqnarray}
where
\begin{eqnarray}
\bar\xi_k=\xi_k\cosh r_k-e^{i\varphi}\xi_k^*\sinh r_k\,.
\end{eqnarray}

By using the above relation, the coherent state is expressed as
\begin{eqnarray}
|\xi_k\rangle_{\rm I}=\hat{D}^{\rm I}\left(\xi_k\right)|0\rangle_{\rm I}=\hat{D}^{\rm R}\left(\bar{\xi}_k\right)|0\rangle_{\rm I}\,.
\end{eqnarray}
This can be also expressed as
\begin{eqnarray}
|\xi_k\rangle_{\rm I}=\hat{D}^{\rm R}\left(\bar{\xi}_k\right)\hat{S}\left(\zeta_k\right)|0\rangle_{\rm R}=\hat{S}\left(\zeta_k\right)\hat{D}^{\rm R}\left(\xi_k\right)|0\rangle_{\rm R}=\hat{S}\left(\zeta_k\right)|\xi_k\rangle_{\rm R}\,.
\label{relation3}
\end{eqnarray}
Similarly, we can get
\begin{eqnarray}
|\xi_k\rangle_{\rm E}=\hat{U}\left(\tilde{\zeta}_k\right)|\xi_k\rangle_{\rm I}\,.
\label{relation4}
\end{eqnarray}

\section{Details of A and B}
\label{appC}

\begin{eqnarray}
AO16&=&8\sin ^4\left(\tilde{\phi}-\phi\right)\sinh ^4\tilde{r}\sinh^42r \cosh2r\Bigl(\cosh2r-\cos\left(\phi-2\theta\right)\sinh2r\Bigr)
\label{ao16}
\,.\\
AO14&=&2\sin ^3\left(\tilde{\phi}-\phi\right)\sinh2\tilde{r}\sinh ^2\tilde{r}\sinh ^32r\left(2-\cosh2r+2\cosh4r\right) \nonumber\\
&&+8\sin\left(2\tilde{\phi}-2\phi\right)\sin^2\left(\tilde{\phi} -\phi\right)\sinh^4\tilde{r}\sinh ^42r \cosh2r\nonumber\\
&&-2\sin^3\left(\tilde{\phi}-\phi\right)\cos\left(\phi-2\theta\right)\sinh2\tilde{r}\sinh ^2\tilde{r}\sinh ^42r \left(-1+4\cosh2r\right)\nonumber\\
&&-8\sin^3\left(\tilde{\phi}-\phi\right)\cos\left(\tilde{\phi}-2\theta\right) \sinh ^4\tilde{r}\sinh ^32r \sinh ^2r
\cosh2r\nonumber\\
&&-8\sin^3\left(\tilde{\phi}-\phi\right)\cos\left(\tilde{\phi}-2\phi+2\theta\right)\sinh ^4\tilde{r}\sinh ^32r\cosh2r \cosh ^2r  \nonumber\\
&&+16\sin ^4\left(\tilde{\phi}-\phi\right)\sin\left(\phi-2\theta\right)\sinh ^4\tilde{r}\sinh ^32r \cosh2r\cosh ^2r \nonumber\\
&&-4\sin\left(2\tilde{\phi}-2\phi\right)\sin^2\left(\tilde{\phi}-\phi\right)\cos\left(\phi-2\theta \right)\sinh ^4\tilde{r}\sinh ^52r\,.
\end{eqnarray}
\begin{eqnarray}
AO12&=&\frac{1}{8}\sin^2\left(\tilde{\phi}-\phi\right)\sinh ^2\tilde{r} \sinh ^22r
\left(2+30\cosh2\tilde{r}+25\cosh\left(2\tilde{r}-4r\right)-8 \cosh\left(2\tilde{r}-2r\right)\right.\nonumber\\
&&\quad-32\cosh2r+14\cosh4r-8\cosh\left(2\tilde{r}+2r\right)+25\cosh\left(2\tilde{r}+4r\right)\left.\right)\nonumber\\
&&
+6\sin^2\left(2\tilde{\phi}-2\phi\right)\sinh ^4\tilde{r} \sinh ^22r \cosh ^2r\sinh ^2r
\nonumber\\
&&
+12\sin\left(2\tilde{\phi}-2\phi\right)\sin\left(\tilde{\phi}-\phi\right)\sinh2\tilde{r} \sinh^2\tilde{r}\sinh ^22r\cosh2r \cosh r\sinh r   \nonumber\\
&&+\sin^2\left(\tilde{\phi}-\phi\right)\cos\left(2\tilde{\phi}-2\phi\right)\sinh^4\tilde{r}\sinh^42r\nonumber\\
&&
+\sin\left(2\tilde{\phi}-2\phi\right)\sin\left(\tilde{\phi}-\phi\right)\sinh2\tilde{r} \sinh^2\tilde{r}\sinh ^32r (-1+2\cosh2r) \nonumber\\
&&
-2\sin^2\left(\tilde{\phi}-\phi\right)\cos\left(\phi -2 \theta\right) \sinh ^2\tilde{r} \sinh ^32r\left(-2-\cosh2\tilde{r}+3\cosh\left(2\tilde{r}-2r\right)+2\cosh2r\right.\nonumber\\
&&\quad\left.+3\cosh\left(2\tilde{r}+2r\right)\right)\nonumber\\
&&
-2\sin^2\left(\tilde{\phi}-\phi\right)\cos \left(\tilde{\phi}-2\theta\right) \sinh2\tilde{r}
   \sinh ^2\tilde{r}\sinh ^22r  \sinh ^2r\left(-1+4\cosh2r\right)\nonumber\\
&&
+4\sin ^3\left(\tilde{\phi}-\phi\right)\sin\left(\phi -2\theta\right) \sinh2\tilde{r}\sinh^2\tilde{r}\sinh ^22r\cosh^2r\left(-1+4\cosh2r\right)\nonumber\\
&&
-2\sin^2\left(\tilde{\phi}-\phi\right)\cos (\tilde{\phi}-2 \phi+2\theta  )\sinh2\tilde{r}\sinh^2\tilde{r}\sinh ^22r\cosh^2r\left(-1+4\cosh2r\right)\nonumber\\
&&
-4\sin\left(2\tilde{\phi}-2\phi\right)\sin\left(\tilde{\phi}-\phi\right)\cos\left(\phi-2\theta \right) \sinh2\tilde{r} \sinh ^2\tilde{r}\sinh ^42r \nonumber\\
&&-4\sin\left(2\tilde{\phi}-2\phi\right)\sin\left(\tilde{\phi}-\phi\right)\cos\left(\tilde{\phi}-2\theta \right) \sinh^4\tilde{r}\sinh^32r\sinh^2r \nonumber\\
&&
+8\sin\left(2\tilde{\phi}-2\phi\right)\sin^2\left(\tilde{\phi}-\psi\right)\sin\left(\phi
   -2\theta\right)\sinh ^4\tilde{r}
   \sinh ^32r \cosh ^2r \nonumber\\
&&
-4 \sin\left(2\tilde{\phi}-2\phi\right)\sin\left(\tilde{\phi}-\phi\right)\cos\left(\tilde{\phi}-2\phi+2\theta \right) \sinh ^4\tilde{r}\sinh ^32r \cosh ^2r\,.
\end{eqnarray}
\begin{eqnarray}
AO10&=&2\sin\left(\tilde{\phi}-\phi\right)\sinh2\tilde{r}\sinh2r\left(\frac{1}{4}\sinh^2\tilde{r}\left(7+5\cosh4r\right)+2\cosh^2\tilde{r}\sinh^2r\left(1+2\cosh2r\right)\right)
\nonumber\\
&&+12\sin\left(2\tilde{\phi}-2\phi\right)\cos\left(\tilde{\phi}-\phi\right)\sinh2\tilde{r}\sinh^2\tilde{r}\sinh2r\cosh ^2r\sinh^2r\nonumber\\
&&-\sin\left(2\tilde{\phi}-2\phi\right)\sinh^2\tilde{r}\sinh^22r\left(2+\cosh2\tilde{r}-4\cosh\left(2\tilde{r}-2r\right)-4\cosh2r-4\cosh\left(2\tilde{r}+2r\right)\right)\nonumber\\
&&+\sin\left(\tilde{\phi}-\phi\right)\cos\left(2\tilde{\phi}-2\phi\right)\sinh2\tilde{r}\sinh^2\tilde{r}\sinh^32r\nonumber\\
&&+\sin\left(\tilde{\phi}-\phi\right)\cos\left(\phi -2 \theta\right)\sinh2\tilde{r}\sinh^22r\left(1+\cosh2\tilde{r}-2\cosh\left(2\tilde{r}-2r\right)-2\cosh\left(2\tilde{r}+2r\right)\right)\nonumber\\
&&-2\sin\left(\tilde{\phi}-\phi\right)\cos \left(\tilde{\phi}-2\theta\right)\sinh^2\tilde{r}\sinh2r\sinh^2r\left(-2-\cosh2\tilde{r}+2\cosh\left(2\tilde{r}-2r\right)\right.\nonumber\\
&&\quad\left.+2\cosh2r+2\cosh\left(2\tilde{r}+2r\right)\right) \nonumber\\
&&+4\sin^2\left(\tilde{\phi}-\phi\right)\sin\left(\phi -2 \theta\right)\sinh^2\tilde{r}\sinh2r\cosh ^2r\left(-2-\cosh2\tilde{r}+2\cosh\left(2\tilde{r}-2r\right)\right.\nonumber\\
&&\quad\left.+2\cosh2r+2\cosh\left(2\tilde{r}+2r\right)\right)\nonumber\\
&&-2\sin\left(\tilde{\phi}-\phi\right)\cos\left(\tilde{\phi}-2\phi+2\theta \right)\sinh^2\tilde{r}\sinh2r\cosh ^2r\left(-2-\cosh2\tilde{r}+2\cosh\left(2\tilde{r}-2r\right)\right.\nonumber\\
&&\quad\left.+2\cosh2r+2\cosh\left(2\tilde{r}+2r\right)\right)\nonumber\\
&&-2\sin\left(2\tilde{\phi}-2\phi\right)\cos\left(\phi -2 \theta\right)\sinh^2\tilde{r}\sinh^32r\left(1+2\cosh2\tilde{r}\right)\nonumber\\
&&-4\sin\left(2\tilde{\phi}-2\phi\right)\cos \left(\tilde{\phi}-2\theta\right)\sinh2\tilde{r}\sinh^2\tilde{r}\sinh^22r\sinh^2r\nonumber\\
&&+8\sin\left(\tilde{\phi}-\phi\right)\sin\left(2\tilde{\phi}-2\phi\right)\sin\left(\phi -2 \theta\right)\sinh2\tilde{r}\sinh^2\tilde{r}\sinh^22r\cosh ^2r\nonumber\\
&&-4\sin\left(2\tilde{\phi}-2\phi\right)\cos\left(\tilde{\phi}-2\phi+2\theta \right)\sinh2\tilde{r}\sinh^2\tilde{r}\sinh^22r\cosh ^2r\\
AO8&=&\frac{1}{32}\left(-14+14\cosh4\tilde{r}-16\cosh\left(2\tilde{r}-2r\right)+9\cosh\left(4\tilde{r}-4r\right)+14\cosh4r\right.\nonumber\\
&&\quad\left.-16\cosh\left(2\tilde{r}+2r\right)+9\cosh\left(4\tilde{r}+4r\right)\right)\nonumber\\
&&+6\cos^2\left(\tilde{\phi}-\phi\right)\sinh^22\tilde{r}\sinh^2r\cosh^2r\nonumber\\
&&+\cos\left(\tilde{\phi}-\phi\right)\sinh2\tilde{r}\sinh2r\left(-1+2\cosh\left(2\tilde{r}-2r\right)+2\cosh\left(2\tilde{r}+2r\right)\right)\nonumber\\
&&+\cos\left(2\tilde{\phi}-2\phi\right)\sinh^2\tilde{r}\cosh^2\tilde{r}\sinh^22r\nonumber\\
&&-\cos\left(\phi -2 \theta\right)\cosh2\tilde{r}\sinh2r\left(\cosh\left(2\tilde{r}+2r\right)+2\sinh^2\left(\tilde{r}+r\right)\right)\nonumber\\
&&-\cos\left(\tilde{\phi} -2 \theta\right)\sinh2\tilde{r}\sinh^2r\left(\cosh\left(2\tilde{r}+2r\right)+2\sinh^2\left(\tilde{r}+r\right)\right)\nonumber\\
&&-\cos\left(\tilde{\phi}-\phi\right)\cos\left(\phi -2 \theta\right)\sinh4\tilde{r}\sinh^22r\nonumber\\
&&-2\cos\left(\tilde{\phi}-\phi\right)\cos\left(\tilde{\phi} -2 \theta\right)\sinh^22\tilde{r}\sinh2r\sinh^2r\nonumber\\
&&+2\sin\left(\tilde{\phi}-\phi\right)\sin\left(\phi -2 \theta\right)\sinh2\tilde{r}\cosh^2r\left(\cosh\left(2\tilde{r}+2r\right)+2\sinh^2\left(\tilde{r}+r\right)\right)\nonumber\\
&&+2\sin\left(2\tilde{\phi}-2\phi\right)\sin\left(\phi -2 \theta\right)\sinh^22\tilde{r}\sinh2r\cosh^2r\nonumber\\
&&-\cos\left(\tilde{\phi}-2\phi+2\theta \right)\sinh2\tilde{r}\cosh^2r\left(\cosh\left(2\tilde{r}+2r\right)+2\sinh^2\left(\tilde{r}+r\right)\right)\nonumber\\
&&-2\cos\left(\tilde{\phi}-\phi\right)\cos\left(\tilde{\phi}-2\phi+2\theta \right)\sinh^22\tilde{r}\sinh2r\cosh^2r\,.\nonumber\\
\end{eqnarray}
\begin{eqnarray}
BO16&=&2\sin^4\left(\tilde{\phi}-\phi\right)\sinh ^4\tilde{r}\sinh^42r \cosh^22r\,.\\
BO14&=&\sin^3\left(\tilde{\phi}-\phi\right)\sinh 2\tilde{r}\sinh^2\tilde{r}\sinh^32r\left(1-\cosh2r+\cosh4r\right)\nonumber\\
&&+2\sin\left(2\tilde{\phi}-2\phi\right)\sin^2\left(\tilde{\phi}-\phi\right)\sinh ^4\tilde{r}\sinh^42r\cosh2r\,.\\
BO12&=&\frac{1}{16}\sin^2\left(\tilde{\phi}-\phi\right)\sinh ^2\tilde{r}\sinh^22r\left(10+22\cosh2\tilde{r}+13\cosh\left(2\tilde{r}-4r\right)-8\cosh\left(2\tilde{r}-2r\right)\right)\nonumber\\
&&\quad-32\cosh2r+6\cosh4r-8\cosh\left(2\tilde{r}+2r\right)+13\cosh\left(2\tilde{r}+4r\right)\nonumber\\
&&+\sin\left(2\tilde{\phi}-2\phi\right)\sin\left(\tilde{\phi}-\phi\right)\sinh2\tilde{r}\sinh^2\tilde{r}\sinh^32r\left(-\frac{1}{2}+2\cosh2r\right)\nonumber\\
&&+\sin^2\left(2\tilde{\phi}-2\phi\right)\sinh^4\tilde{r}\sinh^22r\sinh^2r\cosh^2r\nonumber\\
&&+\frac{1}{2}\sin^2\left(\tilde{\phi}-\phi\right)\cos\left(2\tilde{\phi}-2\phi\right)\sinh^4\tilde{r}\sinh^42r\,.\\
BO10&=&\sin\left(\tilde{\phi}-\phi\right)\sinh2\tilde{r}\sinh2r\left(2\cosh^2\tilde{r}\cosh^2r\sinh^2r+\sinh^2\tilde{r}\left(\cosh^4r+\sinh^4r\right)\right.\nonumber\\
&&\quad\left.+2\cosh^2\tilde{r}\sinh^4r+\sinh^2\tilde{r}\sinh^22r\right)\nonumber\\
&&-\frac{1}{2}\sin\left(2\tilde{\phi}-2\phi\right)\sinh^2\tilde{r}\sinh^22r\left(2-\cosh2\tilde{r}-2\cosh\left(2\tilde{r}-2r\right)-2\cosh2r\right.\nonumber\\
&&\quad\left.-2\cosh\left(2\tilde{r}+2r\right)\right)\nonumber\\
&&+2\sin\left(2\tilde{\phi}-2\phi\right)\cos\left(\tilde{\phi}-\phi\right)\sinh2\tilde{r}\sinh^2\tilde{r}\sinh2r\sinh^2r\cosh^2r\nonumber\\
&&+\frac{1}{2}\sin\left(\tilde{\phi}-\phi\right)\cos\left(2\tilde{\phi}-2\phi\right)\sinh2\tilde{r}\sinh^2\tilde{r}\sinh^32r\,.\\
BO8&=&\frac{1}{64}\left(10+6\cosh4\tilde{r}-16\cosh\left(2\tilde{r}-2r\right)+5\cosh\left(4\tilde{r}-4r\right)+6\cosh4r-16\cosh\left(2\tilde{r}+2r\right)\right.\nonumber\\
&&\quad\left.+5\cosh\left(4\tilde{r}+4r\right)\right)\nonumber\\
&&+\frac{1}{2}\cos\left(\tilde{\phi}-\phi\right)\sinh2\tilde{r}\sinh2r\left(\cosh\left(2\tilde{r}-2r\right)+2\sinh^2\left(\tilde{r}-r\right)\right)\nonumber\\
&&+\frac{1}{2}\cos\left(2\tilde{\phi}-2\phi\right)\cosh^2\tilde{r}\sinh^2\tilde{r}\sinh^22r\nonumber\\
&&+\cos^2\left(\tilde{\phi}-\phi\right)\sinh^22\tilde{r}\sinh^2r\cosh^2r\,.
\end{eqnarray}


\begin{thebibliography}{99}

\bibitem{Einstein:1935rr} 
  A.~Einstein, B.~Podolsky and N.~Rosen,
  Phys.\ Rev.\  {\bf 47}, 777 (1935).

\bibitem{Maldacena:2012xp} 
  J.~Maldacena and G.~L.~Pimentel,
  JHEP {\bf 1302}, 038 (2013)
  [arXiv:1210.7244 [hep-th]].

\bibitem{Kanno:2014lma} 
  S.~Kanno, J.~Murugan, J.~P.~Shock and J.~Soda,
  JHEP {\bf 1407}, 072 (2014)
  [arXiv:1404.6815 [hep-th]].

\bibitem{Iizuka:2014rua} 
  N.~Iizuka, T.~Noumi and N.~Ogawa,
  Nucl.\ Phys.\ B {\bf 910}, 23 (2016)
  [arXiv:1404.7487 [hep-th]].

\bibitem{Kanno:2014bma} 
  S.~Kanno, J.~P.~Shock and J.~Soda,
  JCAP {\bf 1503}, 015 (2015)
  [arXiv:1412.2838 [hep-th]].

\bibitem{Kanno:2016gas} 
  S.~Kanno, J.~P.~Shock and J.~Soda,
  Phys.\ Rev.\ D {\bf 94}, no. 12, 125014 (2016)
  [arXiv:1608.02853 [hep-th]].

\bibitem{Kanno:2016qcc} 
  S.~Kanno, M.~Sasaki and T.~Tanaka,
  JHEP {\bf 1703}, 068 (2017)
  [arXiv:1612.08954 [hep-th]].

\bibitem{Choudhury:2017bou} 
  S.~Choudhury and S.~Panda,
  Eur.\ Phys.\ J.\ C {\bf 78}, no. 1, 52 (2018)
  [arXiv:1708.02265 [hep-th]].

\bibitem{Choudhury:2017qyl} 
  S.~Choudhury and S.~Panda,
  arXiv:1712.08299 [hep-th].

\bibitem{Albrecht:2018prr} 
  A.~Albrecht, S.~Kanno and M.~Sasaki,
  Phys.\ Rev.\ D {\bf 97}, no. 8, 083520 (2018)
  [arXiv:1802.08794 [hep-th]].

\bibitem{Abbott:2016blz} 
  B.~P.~Abbott {\it et al.} [LIGO Scientific and Virgo Collaborations],
  Phys.\ Rev.\ Lett.\  {\bf 116}, no. 6, 061102 (2016)
  [arXiv:1602.03837 [gr-qc]].

\bibitem{Kawamura:2011zz} 
  S.~Kawamura {\it et al.},
  Class.\ Quant.\ Grav.\  {\bf 28}, 094011 (2011).

\bibitem{AmaroSeoane:2012km} 
  P.~Amaro-Seoane {\it et al.},
  GW Notes {\bf 6}, 4 (2013)
  [arXiv:1201.3621 [astro-ph.CO]].

\bibitem{Agarwal:2012}
Girish.~S.~Agarwal,
Quantum Optics, Cambridge University Press (2012)

\bibitem{Kanno:2018cuk} 
  S.~Kanno and J.~Soda,
  Phys.\ Rev.\ D {\bf 99}, no. 8, 084010 (2019)
  [arXiv:1810.07604 [hep-th]].

\bibitem{Koh:2004ez} 
  S.~Koh, S.~P.~Kim and D.~J.~Song,
  JHEP {\bf 0412}, 060 (2004)
  [gr-qc/0402065].

\bibitem{Kundu:2011sg} 
  S.~Kundu,
  JCAP {\bf 1202}, 005 (2012)
  [arXiv:1110.4688 [astro-ph.CO]].

\bibitem{Glauber:1963fi} 
  R.~J.~Glauber,
  Phys.\ Rev.\  {\bf 130}, 2529 (1963).

\bibitem{Watanabe:2009ct} 
  M.~a.~Watanabe, S.~Kanno and J.~Soda,
  Phys.\ Rev.\ Lett.\  {\bf 102}, 191302 (2009)
  [arXiv:0902.2833 [hep-th]].
  
\bibitem{Soda:2012zm} 
  J.~Soda,
  Class.\ Quant.\ Grav.\  {\bf 29}, 083001 (2012)
  [arXiv:1201.6434 [hep-th]].

\bibitem{Maleknejad:2012fw} 
  A.~Maleknejad, M.~M.~Sheikh-Jabbari and J.~Soda,
  Phys.\ Rept.\  {\bf 528}, 161 (2013)
  [arXiv:1212.2921 [hep-th]].

\bibitem{Barnaby:2010vf} 
  N.~Barnaby and M.~Peloso,
  Phys.\ Rev.\ Lett.\  {\bf 106}, 181301 (2011)
  [arXiv:1011.1500 [hep-ph]].

\bibitem{Barnaby:2011vw} 
  N.~Barnaby, R.~Namba and M.~Peloso,
  JCAP {\bf 1104}, 009 (2011)
  [arXiv:1102.4333 [astro-ph.CO]].
  
\bibitem{Cook:2011hg} 
  J.~L.~Cook and L.~Sorbo,
  Phys.\ Rev.\ D {\bf 85}, 023534 (2012)
  Erratum: [Phys.\ Rev.\ D {\bf 86}, 069901 (2012)]
  [arXiv:1109.0022 [astro-ph.CO]].
  
\bibitem{Anber:2012du}
  M.~M.~Anber and L.~Sorbo,
  Phys.\ Rev.\ D {\bf 85} (2012) 123537
  [arXiv:1203.5849 [astro-ph.CO]].
  
\bibitem{Barnaby:2011qe} 
  N.~Barnaby, E.~Pajer and M.~Peloso,
  Phys.\ Rev.\ D {\bf 85}, 023525 (2012)
  [arXiv:1110.3327 [astro-ph.CO]].

\bibitem{Choi:2015wva} 
  K.~Choi, K.~Y.~Choi, H.~Kim and C.~S.~Shin,
  JCAP {\bf 1510}, no. 10, 046 (2015)
  [arXiv:1507.04977 [astro-ph.CO]].

\bibitem{Ito:2016aai} 
  A.~Ito and J.~Soda,
  JCAP {\bf 1604}, no. 04, 035 (2016)
  [arXiv:1603.00602 [hep-th]].

\bibitem{Giovannini:2019bfw} 
  M.~Giovannini,
  arXiv:1902.11075 [hep-th].

\bibitem{Giovannini:2019ehc} 
  M.~Giovannini,
  arXiv:1903.03796 [gr-qc].

\bibitem{HanburyBrown:1956bqd} 
  R.~Hanbury Brown and R.~Q.~Twiss,
  Nature {\bf 178}, 1046 (1956).

\bibitem{Brown:1956zza} 
  R.~H.~Brown and R.~Q.~Twiss,
  Nature {\bf 177}, 27 (1956).

\bibitem{Giovannini:2010xg} 
  M.~Giovannini,
  Phys.\ Rev.\ D {\bf 83}, 023515 (2011)
  [arXiv:1011.1673 [astro-ph.CO]].

\bibitem{Giovannini:2016esa} 
  M.~Giovannini,
  Class.\ Quant.\ Grav.\  {\bf 34}, no. 3, 035019 (2017)
  [arXiv:1608.05843 [hep-th]].

\bibitem{Giovannini:2017uty} 
  M.~Giovannini,
  Mod.\ Phys.\ Lett.\ A {\bf 32}, no. 35, 1750191 (2017)
  [arXiv:1709.00914 [gr-qc]].

\bibitem{Chen:2017cgw} 
  J.~W.~Chen, S.~H.~Dai, D.~Maity, S.~Sun and Y.~L.~Zhang,
  arXiv:1701.03437 [quant-ph].
 


  
\end{thebibliography}
\end{document}